\documentclass[12pt]{article}
\pdfoutput=1

\usepackage[bulletsep]{collref}
\usepackage{amssymb,graphicx}
\usepackage[intlimits]{amsmath}
\usepackage{bbm}
\usepackage[small]{subfigure}

\usepackage{MnSymbol}

\makeatletter \@addtoreset{equation}{section} \makeatother

\makeatletter
\let\old@startsection=\@startsection
\let\oldl@section=\l@section
\renewcommand{\@startsection}[6]{\old@startsection{#1}{#2}{#3}{#4}{#5}{#6\mathversion{bold}}}
\renewcommand{\l@section}[2]{\oldl@section{\mathversion{bold}#1}{#2}}
\makeatother

\makeatletter
\let\old@makecaption=\@makecaption
\def\@makecaption{\small\old@makecaption}
\makeatother

\renewcommand{\leq}{\leqslant}
\renewcommand{\geq}{\geqslant}

\newcommand{\la}{\lambda}

\newcommand{\bC}{\mathbb C}
\newcommand{\bCP}{\mathbb {CP}}
\newcommand{\bR}{\mathbb R}

\newcommand {\cP} {H}

\newcommand{\si}{\sigma}

\newcommand{\bB}{\mathcal B}

\newcommand{\ep}{\boldsymbol{\varepsilon}}
\newcommand{\et}{\boldsymbol{\eta}}
\newcommand{\nd}{\boldsymbol{\nu}}

\begin{document}


\begin{flushright}\footnotesize
\texttt{NORDITA-2016-19} \\
\texttt{UUITP-02/16}
\vspace{0.6cm}
\end{flushright}

\renewcommand{\thefootnote}{\fnsymbol{footnote}}
\setcounter{footnote}{0}

\begin{center}
{\Large\textbf{\mathversion{bold} Level Crossing in Random Matrices: \\I.
 Random perturbation of a fixed matrix}
\par}

\vspace{0.8cm}

\textrm{B.~Shapiro$^{1}$ and
K.~Zarembo$^{2,3}$}
\vspace{4mm}

\textit{${}^1$Department of Mathematics, Stockholm University, SE-106 91 Stockholm, Sweden}\\
\textit{${}^2$Nordita,  Stockholm University and KTH Royal Institute of Technology,
Roslagstullsbacken 23, SE-106 91 Stockholm, Sweden}\\
\textit{${}^3$Department of Physics and Astronomy, Uppsala University\\
SE-751 08 Uppsala, Sweden}\\
\vspace{0.2cm}
\texttt{shapiro@math.su.se, zarembo@nordita.org}

\vspace{3mm}


\par\vspace{1cm}

\textbf{Abstract} \vspace{3mm}

\begin{minipage}{13cm}
We consider level crossing in a matrix family $H=H_0+\lambda V$ where $H_0$ is a fixed $N\times N$ matrix and $V$ belongs to one of the standard Gaussian random matrix ensembles. We study the probability distribution of level crossing points in the complex plane of $\lambda $, for which we obtain a number of exact, asymptotic and approximate formulas.
\end{minipage}

\end{center}

\vspace{0.5cm}


\newpage
\setcounter{page}{1}
\renewcommand{\thefootnote}{\arabic{footnote}}
\setcounter{footnote}{0}

\section{Introduction}

Analysis of the dependence  of the spectrum  on a perturbative parameter $\la$  in  linear families   
\begin{equation}\label{ham} 
\cP=H_0+\la V,
\end{equation}
  is a typical problem both   in physics and mathematics, see e.g. the  treatise \cite{Ka}.  Here $H_0$ is an initial linear operator,  $V$ is a perturbing linear operator, and $\la$ is  a real/complex-valued  parameter.

\medskip
 In many  concrete situations  $H_0$ and $V$ are self-adjoint and $\la$ is real, which typically leads to the conclusion that,  for all real values of $\la$,  the spectrum is real  and simple. Without special symmetry reasons the eigenvalues of a Hermitian matrix do not cross, as a consequence of the famous von~Neumann-Wigner eigenvalue repulsion \cite{vonNeumann:1929:VEA}. The spectrum of $H$ for real $\lambda $ is consequently classified by the energy levels of the original, unperturbed Hamiltonian $H_0$. 
 
 Since the late 60's,  motivated by a number of fascinating observations   by C.~M.~Bender and T.~T.~Wu  \cite{BW}, physicists and  mathematicians    started considering various cases  where $H_0$ and $V$ are, for example,  self-adjoint while $\la$ is  complex-valued.  The level crossings occur upon analytic continuation of  $\lambda $ into the complex plane, where an intricate pattern of level permutations occurs due to monodromy at each of the branch points. The positions and monodromy of the  level crossings constitute an important piece of information about the spectral data for the linear family  (\ref{ham}) and its analytic structure. They determine, for instance, the accuracy of perturbative series in $\lambda $. 

\medskip
  One of the basic  questions in such a study  posed by C.~M.~Bender and T.~T.~Wu is whether it is possible to  interchange two arbitrary chosen real eigenvalues $E_i$ and $E_j$ of \eqref{ham} corresponding to some fixed real value  $\la_0,$ by allowing    $\la$ to move in the complex plane along some closed loop which starts and ends at  $\la_0$.  The latter question can be rephrased as the connectivity of  the corresponding spectral surface $S_\cP$,   where the 
$S_\cP\subset \bC^2$ is the set of all pairs $(E,\la)\in \bC^2,$ $E$ being an eigenvalue of $\cP$ with a given value of parameter $\la$. In  several interesting situations it is proven that $S_\cP$ is a complex-analytic curve in $\bC^2$ given as the zero locus of an appropriate entire function in two variables called the {\it spectral determinant}.  Important results in this direction were recently obtained in e.g. \cite{EG} and \cite{AG}. (For families of finite-dimensional matrices,  $S_\cP$ is an algebraic curve given by the  spectral equation~\eqref{eq:curve}   below.)

\medskip
For large classes of linear families \eqref{ham} of infinite-dimensional linear operators,   there exists a discrete level crossing set  $\bB_\cP\subset\bC$ consisting of all values of $\la$ for which the spectrum of  \eqref{ham} is not simple. 
(In physics literature such points are often referred to as  ``exceptional points"; the latter term was coined by T.Kato in \cite{Ka}.)  In particular, for generic  families \eqref{ham}  of  $n\times n$-matrices,  the level crossing  set $\bB_\cP\subset \bC$ consists of  $n(n-1)$ distinct exceptional points. 

For many concrete families of linear differential or matrix  operators,  it is highly desirable  to get  information about their level crossing sets $\bB_\cP$ as well as about the monodromy of the eigenvalues, when the parameter $\lambda $ traverses  different  closed loops  in $\bC\setminus \bB_\cP$.  Unfortunately the latter problem (especially its monodromy part) seems, in general, to be quite hard,  see  examples in e.g. \cite{ShTaSe}, \cite{ShTaQu}. 

\medskip
Notice that for matrix families with Hermitian $H_0$ and $V,$  studies of the corresponding spectral surfaces and their branch (=exceptional) points are related to the famous Lax conjecture, see \cite{Lax} and determinantal representations of polynomials, see e.g. \cite{PlVi}.  It turns out that one can explicitly characterise  the class of real spectral determinants  = real algebraic curves given by the equation 
\begin{equation}\label{eq:curve}
\chi(\la,E):=\det (H_0+\la V+E I)=\det (\cP+E I)=0,
\end{equation}  
with arbitrary Hermitian $H_0$ and $V$ of some size $n$. 
 For complex-valued square matrices $H_0$ and $V$ of a given size $n,$ it was already shown by A.~C.~Dixon, \cite {Di}  in 1902  that any  plane complex algebraic curve of degree $n$  can be represented by \eqref{eq:curve}.   He also found  how many different determinantal representations there exist for a generic plane curve of degree $n$. 

\medskip 
Observe also that level crossing  sets $\bB_\cP\subset \bC$  which can appear as the sets of branch points of complex plane curves of degree $n$ (or, equivalently, of representations \eqref{ham}) contain $n(n-1)$ points but depend only on $\binom {n+2}{2}-4=\frac{n^2+3n-6}{2}$  parameters.  This  means that,  starting with $n=4,$ there exist (quite complicated) relations among the branch points, see \cite{OnSh}. In the first non-trivial case $n=4$   there exists one relation on the 12  points in the level crossing set, i.e., these configurations of 12 points   form a  hypersurface in $\bCP^{12}$ which was considered in  \cite {Va}. In particular, in \cite {Va} it was shown  that the degree of this hypersurface equals  $3762$.

\medskip
Energy level repulsion is ubiquitous in quantum mechanics  \cite {STVB}. The level crossing, when it happens at real values of the coupling constants, is a powerful diagnostic of hidden symmetries \cite{YAS}. Level crossings away from the real line, which always occur, signal, in many cases, the change of regime or near-resonance behavior (e.g.  \cite{BR}).

\medskip

In the present paper, instead of looking at concrete families \eqref{ham}, we will utilize the point of view of random matrix theory. 
Namely, we will study spectra and level crossings in  \eqref{ham}, assuming that $H_0$ is a given fixed matrix, while $V$ is a random matrix with known distribution. This can be regarded as a crude model for a quantum-mechanical system subject to a   random perturbation. 

Since in our approach the matrices at hand are random, it is appropriate to talk about statistics of level crossings, and at least two important questions can be posed in this setup. The first one is the distribution of  level crossings in the complex plane of  $\lambda$. To the best of our knowledge, for the first time this question was  addressed in  \cite{zirnbauer1983destruction} in the context of quantum chaos for nuclear spectra, similar problem was addressed in  \cite{BEDPSW} in conjunction
of topologically protected Andreev level crossings in Josephson junctions.
The second, and a more  difficult question is how to describe statistical properties of the spectral monodromy.  We leave aside the study of monodromy for a future work and concentrate on the statistics of level crossing in this paper.

\section{Level Crossing in random environment}

We assume that $V$ and $H_0$ are $N\times N$ matrices, $H_0$ is  fixed  and $V$ is randomly chosen from one of the standard random matrix ensembles
\cite{mehta2004random,anderson2010an}. More concretely, we discuss below the four cases of Gaussian unitary, Gaussian orthogonal, real and complex Gaussian ensembles. (The remaining classical case of Gaussian symplectic ensemble exhibits Kramers degeneracy of spectrum and does not  seem to be suitable for our study both from the theoretical and numerical perspectives.)\footnote{The case when $H_0$ is also random  will be considered in the sequel \cite{ShZa2,ShZa3}.}
  
    Without loss of generality  $H_0$ can be taken diagonal:
\begin{equation}
 H_0=\mathop{\mathrm{diag}}\left(E_1,\ldots ,E_N\right).
\end{equation}
We additionally assume that $E_1,\ldots ,E_N$ are pairwise distinct.  For any given perturbation $V\neq 0$, the eigenvalues of the matrix (\ref{ham}) collide pairwise at $N(N-1)$ generically distinct complex values of the coupling parameter $\lambda$.  The probability distribution of the matrix $V$ induces a statistical distribution of the level-crossing points in the complex plane of $\lambda $.  When $V$ belongs to the Gaussian Orthogonal Ensemble (GOE), this distribution was calculated analytically for $N=2$  and  was studied numerically for larger $N$   in \cite{zirnbauer1983destruction}.  

In the general case of $N\times N$ matrices,  the level-crossing condition is equivalent to the vanishing of the discriminant of the characteristic equation of the  matrix $H$ in (\ref{ham}) which gives a polynomial equation of degree  $N(N-1)$ in the complex variable $\lambda $. For $N=2$ this equation is quadratic  and the PDF (probability density function) of level crossings for $2\times 2$ matrices can be calculated in the closed form.
The discriminantal equation becomes increasingly complicated with growing $N$, and already for $N=3$ the formulas are so cumbersome that we will not present them here. Instead we will use the  explicit solution for the $2\times 2$ case to obtain both asymptotic and approximate results for matrices of arbitrary size. 

The heuristics behind this approach is that near a level-crossing point, the problem always reduces to the two-level interaction. Our arguments are similar in spirit to the textbook derivation of level repulsion  from  the $2\times 2$ secular perturbation theory near a would-be crossing point  \cite{landau1991quantum}.

More concretely, we calculate the exact asymptotics of the level-crossing PDF  at weak coupling (i.e., for small $|\lambda|$) for any $N$.  Additionally, we propose an approximation for the level-crossing PDF under a heuristic assumption that collisions of different pairs of eigenvalues are statistically independent events. Quite surprisingly, this simple-minded approximation is extremely accurate and agrees very well with the actual level-crossing PDF which we confirm by extensive numerical simulations.

\section{Gaussian Unitary Ensemble}

The case that we study most thoroughly is the one of the Gaussian Unitary Ensemble, $GUE_N$. Then 
$V$ is a random  $N\times N$ Hermitian matrix whose entries  have Gaussian statistical distribution. The probability measure is  given by
\begin{equation}\label{Gauss}
 d\mathcal{P}(V)=\left(2\pi \sigma^2 \right)^{-\frac{N^2}{2}}\,{\rm e}\,^{-\frac{1}{2\sigma ^2}\,\mathop{\mathrm{tr}}V^2}\prod_{1\le i\le j\le N}^{}dV_{ij},
\end{equation}
where $\sigma^2$ is the variance.

\subsection{$2\times 2$ case} 

We start with the simplest case of $2\times 2$ matrices. Then $H_0=\mathop{\mathrm{diag}}\left(E_1, E_2\right)$ and $V\in GUE_2$.  The crossing probability depends only on the difference 
\begin{equation}
 \Delta =E_2-E_1.
\end{equation}

It is convenient to expand all matrices at hand in the basis $\left\{\mathbbm{1},\boldsymbol{\sigma }\right\}$, where $\boldsymbol{\sigma }=(\sigma_1,\sigma_2,\sigma_3)$ is  the usual triple of Pauli matrices:
\begin{equation}
 \cP=h_\mathbbm{1} \mathbbm{1}+\mathbf{h}\cdot \boldsymbol{\sigma },
\end{equation}
where $h_\mathbbm{1} \in \bC$ and $\mathbf{h}\in \bC^3$. 
Observe that 
$$
\det \cP=h_\mathbbm{1}^2-\mathbf{h}^2,
$$
where  $\mathbf{w}^2$ stands for the sum of squares of components of vector $\mathbf{w}$.
Therefore the characteristic equation for the matrix $\cP$ reads
\begin{equation}
 \left(h_\mathbbm{1}-E\right)^2=\mathbf{h}^2,
\end{equation}
and the condition for level crossing (i.e., the coincidence of the two eigenvalues of $H$) is simply
\begin{equation}\label{h2=0}
 \mathbf{h}^2=0.
\end{equation}

Now expanding $H_0$ and $V$ from (\ref{ham}) in the Pauli matrices:
\begin{equation}\label{Pauliexp}
 H_0=\varepsilon _\mathbbm{1}\mathbbm{1}+\boldsymbol{\varepsilon }\cdot \boldsymbol{\sigma },
 \qquad 
 V=v_\mathbbm{1}\mathbbm{1}+\mathbf{v}\cdot \boldsymbol{\sigma },
\end{equation}
we see that the level crossing happens when $\lambda  $ satisfies the quadratic equation
\begin{equation}\label{eq:quadr}
 \lambda^2\mathbf{v}^2+2\lambda  \mathbf{v}\cdot \boldsymbol{\varepsilon }+\boldsymbol{\varepsilon }^2=0.
\end{equation}
The vectors $\boldsymbol{\varepsilon }$ and $\mathbf{v}$ have real components since $H_0$ and $V$ are Hermitian matrices.
One can notice that $\varepsilon:= |\boldsymbol{\varepsilon }|=(E_2-E_1)/2=\Delta/2 $. Denoting the angle between $\mathbf{v}$ and $\boldsymbol{\varepsilon}$ by $\theta $ and using $\boldsymbol{\varepsilon}\cdot \mathbf{v}=\varepsilon v\cos\theta $, where $v:=|\mathbf{v}|$,  we can  
explicitly solve equation~\eqref{eq:quadr}:
\begin{equation}\label{abara}
 \lambda =-\frac{\Delta }{2v}\,\,{\rm e}\,^{ i\theta  },\qquad 
 \bar{\lambda }=-\frac{\Delta }{2v}\,\,{\rm e}\,^{ -i\theta  }.
\end{equation}

The level-crossing condition thus explicitly expresses $\lambda $ in terms of  the fixed eigenvalue difference $\Delta $ of $H_0$ and the random variables $v$ and $\theta$. The problem reduces to calculating the
probability distributions for $v$ and $\theta$  using the known  PDF for the matrix elements of $V$. Integrating out $v_\mathbbm{1}$ and expressing the probability measure in the coordinates  $v=|\mathbf{v}|$ and $\theta $, we get:
\begin{equation}
 d\mathcal{P}(v,\theta )=\frac{2v^2\sin\theta }{\sqrt{\pi }\,\sigma ^3}\,
 \,{\rm e}\,^{-\frac{v^2}{\sigma ^2}}dvd\theta.
\end{equation}
The   level-crossing PDF  is now given by the Jacobian of the transformation from the variables $(v,\theta )$ to the position of the level crossing $(\lambda ,\bar{\lambda })$ in the complex plane. Using (\ref{abara}) we obtain:
\begin{equation}\label{eq:cond}
 d\mathcal{P}_{U_2}(\lambda ,\bar{\lambda })
 =\frac{\Delta ^3}{16\sqrt{\pi }\,\sigma ^3}\,\,\frac{\left|\mathop{\mathrm{Im}}\lambda \right|}{\left|\lambda \right|^6}\,\,{\rm e}\,^{-\frac{\Delta ^2}{4\sigma ^2\left|\lambda \right|^2}}d\lambda d\bar{\lambda }.
\end{equation}
An extra factor of $1/2$ arises because there are two level-crossing points for any realization of the random matrix $V$, and we normalize the probability to one.

\begin{figure}[t]
\begin{center}
 \centerline{\includegraphics[width=8cm]{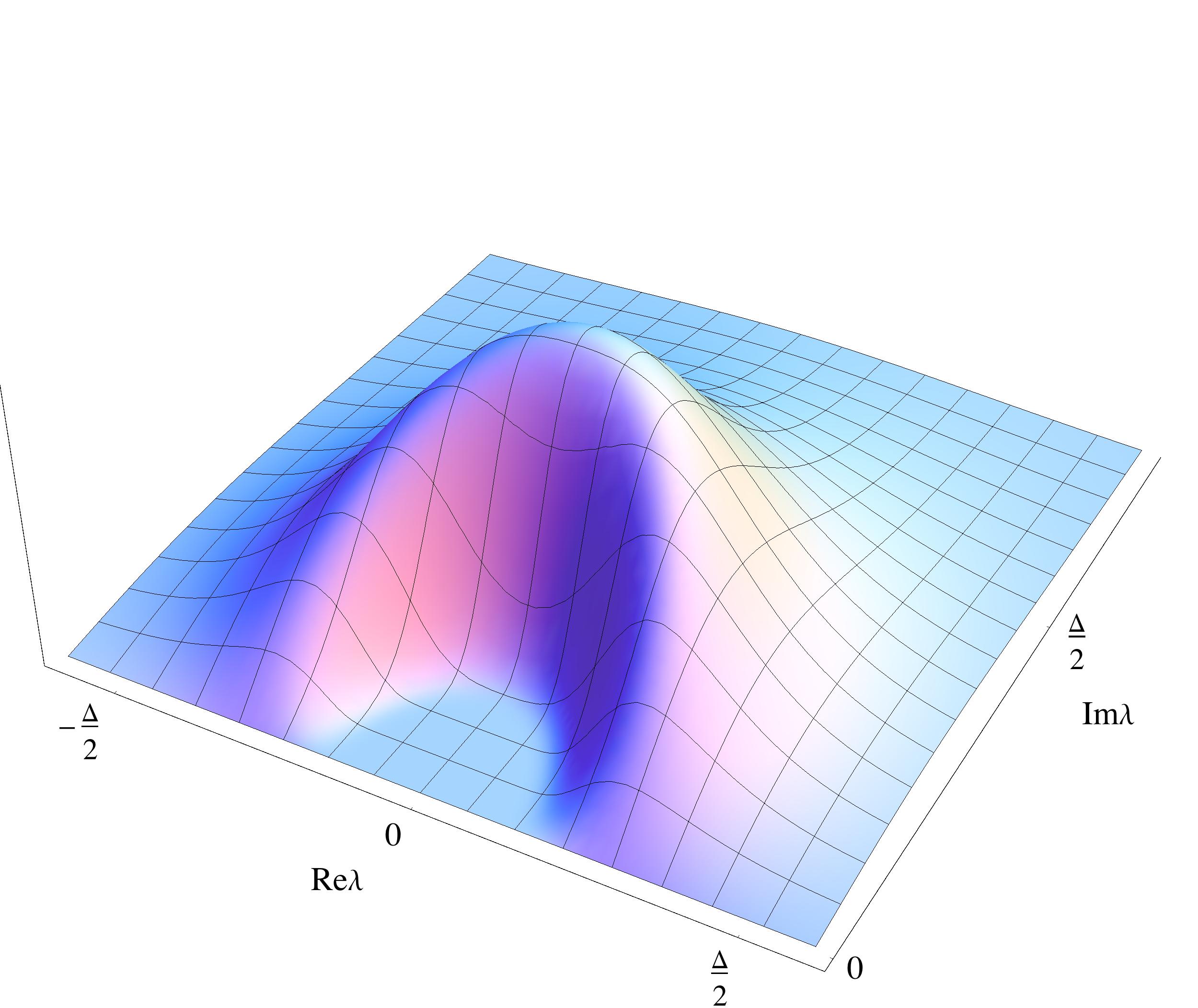}}
\caption{\label{pd}\small The level-crossing probability density (for $\sigma =1$).}
\end{center}
\end{figure}

The level-crossing probability density is shown in fig.~\ref{pd}. The density vanishes on the real line, when $\mathop{\mathrm{Im}}\lambda =0$, as a manifestation of the level repulsion for Hermitian matrices.

Equation~\eqref{eq:cond} takes a more elegant form in the variables
\begin{equation}\label{p+iq}
 \frac{1}{\lambda }=p+iq.
\end{equation}
In particular, the cumulative probability to find the level-crossing point at $q^2>x$ obeys the standard  Poisson distribution:
\begin{equation}\label{Poisson}
  \mathcal{P}_{U_2}(q^2>x)
 =\,{\rm e}\,^{-\frac{\Delta ^2x}{4\sigma ^2}}.
\end{equation}

\subsection{Weak-coupling asymptotics}\label{weak-c-sec}

We will not be able to calculate exactly  the level-crossing PDF  for matrices of a larger size, but we will find some asymptotic and approximate results which are valid for any $N$. Before we proceed,  it is instructive to take another  look at the  formulas for $N=2$. 
When $|\lambda| $ is small, the probability of level crossing (\ref{eq:cond}) is exponentially small, and it is easy to understand why.
A perturbation of strength  $\lambda V\sim \Delta$ is necessary to close the gap of size $\Delta $. Since $|\lambda|$ is small, such a perturbation occurs with an exponentially small probability $\mathcal{P}\sim \exp(-V^2/\sigma ^2)\sim \exp(-\Delta ^2/\sigma ^2|\lambda|^2)$. Clearly, the same heuristics applies to matrices of arbitrary size, and obviously the easiest gap to close is the smallest  gap in the spectrum of $H_0$  (assumed unique). The weak-coupling  asymptotics of the level-crossing probability therefore is determined by the two closest eigenvalues of $H_0$. As before, let us  denote the smallest gap in the spectrum of $H_0$ by $\Delta $, and assume without the loss of generality that $\Delta$ occurs between $E_1$ and $E_2$.

The matrix $H$ then has the  form: 
\begin{equation}
 H=\begin{pmatrix}
 \frac{\Delta }{2}\,\sigma _3+\lambda v_\mathbbm{1}\,\mathbbm{1}+\lambda \mathbf{v}\cdot\boldsymbol{\sigma }  &  \lambda F\\ 
  \lambda F^\dagger  & A+\lambda B \\ 
 \end{pmatrix}.
\end{equation}
  Here $A=\mathop{\mathrm{diag}}(\tilde E _3,\ldots ,\tilde E _N)$ with\footnote{The shift by $\frac{E_1+E_2}{2}$ is necessary to place the two closets levels symmetrically with respect to zero. The level-crossing PDF depends only on the relative distances between the eigenvalues of $H_0$ and is invariant under the shift of the whole spectrum by a common constant.}
\begin{equation}\label{shifetspectrum}
 \tilde E _k=E_k-\frac{E_1+E_2}{2}\,,
\end{equation}
  $B$ is a random $(N-2)\times (N-2)$ Hermitian matrix, $F$ is a random  $2\times (N-2)$ complex matrix with independent Gaussian entries, and, finally,  $(v_\mathbbm{1},\mathbf{v})$ is a random vector, the same as in the above discussion of the $2\times 2$ case. 
    
  In the zeroth-order approximation we can neglect both $\lambda F$ and $\lambda B$ terms. On the contrary we cannot assume that $\lambda \mathbf{v}$ is small. The fluctuation in $\mathbf{v}$ must  be large, $O(1/\lambda )$, in order to close the gap $\Delta $. Of course such large fluctuation occurs with an exponentially small probability.
  In this approximation the $2\times 2$ subsystem of the first two levels decouples, yielding 
the level-crossing PDF which is exactly the same as in the $2\times 2$ case. 
 Let us compute the next order in $\lambda $. As we shall see this affects the overall normalization factor. 
 
 It is clear from~(\ref{h2=0}) that the two eigenvalues of the $2\times 2$ block we are concentrating upon cross at zero. In order to take into account the feedback from the "spectator" levels, we need to solve the equation
\begin{equation}
 \begin{pmatrix}
 \frac{\Delta }{2}\,\sigma _3+\lambda v_\mathbbm{1}\,\mathbbm{1}+\lambda \mathbf{v}\cdot\boldsymbol{\sigma }  &  \lambda F\\ 
  \lambda F^\dagger  & A+\lambda B \\ 
 \end{pmatrix}\begin{pmatrix}
  \psi  \\ 
  \chi  \\ 
 \end{pmatrix}=0.
\end{equation}
Upon excluding $\chi $ via the relation:
\begin{equation}
 \chi =-\lambda \left(A+\lambda B\right)^{-1}F^\dagger \psi \approx -\lambda 
 A^{-1}F^\dagger \psi ,
\end{equation}
 we arrive at an effective two-dimensional problem with the $2\times 2$ matrix
\begin{equation}
 H_{\rm eff}= \frac{\Delta }{2}\,\sigma _3+\lambda v_\mathbbm{1}\,\mathbbm{1}+\lambda \mathbf{v}\cdot\boldsymbol{\sigma } -\lambda ^2FA^{-1}F^\dagger .
\end{equation}
The last term is the next-order correction we were looking for.

Consider now a complex vector in $\bC^3$ given by: 
\begin{equation}
 \mathbf{a}+i\mathbf{b}:=\frac{\Delta }{2\lambda }\mathbf{e}_3-\frac{\lambda }{2}\,\mathop{\mathrm{tr}}(\boldsymbol{\sigma }FA^{-1}F^\dagger),
\end{equation}
where $\mathbf{a}$ and $\mathbf{b}$ are real vectors and $\mathbf{e}_3=(0,0,1)^t$. The last expression $\mathrm{tr} (\boldsymbol{\sigma }FA^{-1}F^\dagger)$ appearing here should be understood as a vector in $\bC^3$ obtained by 
multiplication of the triple of Pauli matrices $\boldsymbol{\sigma }=(\si_1,\si_2,\si_3)$ by the $2\times 2$ matrix $FA^{-1}F^\dagger$ and taking  trace of the product.

Since $H_{\rm eff}=\lambda (\mathbf{v}+\mathbf{a}+i\mathbf{b})\cdot\boldsymbol{\sigma }+\lambda v_\mathbbm{1}\mathbbm{1}$, 
the conditions that $H_{\rm eff}$ has coinciding eigenvalues is given by:
\begin{equation}\label{v+a+ib}
 \left(\mathbf{v}+\mathbf{a}+i\mathbf{b}\right)^2=0,
\end{equation}
or, equivalently, by: 
$$
\begin{cases}
 \left(\mathbf{v}+\mathbf{a}\right)^2=\mathbf{b}^2
\nonumber \\
\left(\mathbf{v}+\mathbf{a}\right)\cdot \mathbf{b}=0.
\end{cases}
$$
The solutions to these equations form a circle:
\begin{equation}\label{v=bn-a}
 \mathbf{v}=b\mathbf{n}-\mathbf{a},
\end{equation}
where $b:=|\mathbf{b}|$ and $\mathbf{n}$ is a unit vector perpendicular to $\mathbf{b}$, i.e., 
\begin{equation}\label{bn-eq}
 \mathbf{b}\cdot \mathbf{n}=0,\qquad \mathbf{n}^2=1.
\end{equation}
The level-crossing probability is the length of this circle,  measured with respect to the probability density
\begin{equation}\label{vectorPDF}
 d\mathcal{P}(\mathbf{v})=\frac{{{\rm e}^{-\frac{v^2}{\sigma ^2}}\,\,d^3\mathbf{v}}} {{{\pi^{\frac{3}{2}}} \sigma ^3}}
\end{equation}
and averaged over the $2\times (N-2)$ random matrix $F$ which has independent complex Gaussian entries.

Substituting $ \lambda =x+iy$ and using the standard properties of the trace, 
the two vectors $\mathbf{a}$ and $\mathbf{b}$ become:
\begin{equation} \label{ab-to-lineraorder}
\begin{cases}
 \mathbf{a}=\frac{\Delta x}{2|\lambda |^2}\,\mathbf{e}_3-\frac{x}{2}\,\mathop{\mathrm{tr}}(A^{-1}F^\dagger \boldsymbol{\sigma }F), \\
 \mathbf{b}=-\frac{\Delta y}{2|\lambda |^2}\,\mathbf{e}_3-\frac{y}{2}\,\mathop{\mathrm{tr}}(A^{-1}F^\dagger \boldsymbol{\sigma }F). 
\end{cases}
\end{equation}
The first terms in the right-hand side of \eqref{ab-to-lineraorder} are of order $O(1/\lambda )$, while the second ones are of order $O(\lambda )$ and  can be neglected to the first approximation.  The real vectors $\mathbf{a}$ and $\mathbf{b}$ are then collinear, and (\ref{v=bn-a}) is a circle of radius $b$ in the lateral plane shifted by distance $a:=|\mathbf{a}|$ in the $\mathbf{e}_3$ direction.  The probability measure (\ref{vectorPDF}) is constant on this circle, and the level-crossing PDF  is obtained by changing variables 
$d^3\mathbf{v}\rightarrow J\,dl d\lambda d\bar{\lambda }$, where $dl $ is the line element on the circle and $J\propto |\lambda |^{-4}$ is the Jacobian. The integration over $dl$ gives the length of the circle, proportional to $|y|/|\lambda |^2$, which altogether results in the  equation~(\ref{eq:cond}) derived above.

When we take into account the correction term, the circle gets slightly tilted. The correction to the prefactor in (\ref{eq:cond}) will be small with perturbation, and can be safely neglected, while the correction to the exponent is of order one, and has to be taken into account. The variation $\delta v^2$ up to the linear order in  $\delta \mathbf{a}$ and $\delta \mathbf{b}$ is given by: 
\begin{equation}
 \delta v^2=2\mathbf{a}\cdot \delta \mathbf{a}+2\mathbf{b}\cdot \delta \mathbf{b}-2b\,\delta \mathbf{a}\cdot \mathbf{n}-2b\,\mathbf{a\cdot \delta n},
\end{equation}
where by $\delta \mathbf{a}$ and $\delta \mathbf{b}$ we denote the second terms in (\ref{ab-to-lineraorder}).
Observe that 
we do not need to include the variation of $b$, because $\mathbf{a}\cdot \mathbf{n}=0$ to the leading order. Linearizing  condition (\ref{bn-eq}), we find:
\begin{equation}
 \delta \mathbf{n}=-\left(\delta \mathbf{b}\cdot \mathbf{n}\right)\,\frac{\mathbf{b}}{b^2}\,,
\end{equation}
and substituting the explicit expressions for $\delta \mathbf{a}$ and $\delta \mathbf{b}$ from (\ref{ab-to-lineraorder}) into the above variation of $v^2$, we obtain:
\begin{equation}
 \delta v^2=\frac{\Delta }{2|\lambda |^2}\,\mathop{\mathrm{tr}}\left\{A^{-1}F^\dagger \left[\left(y^2-x^2\right)\sigma _3+2x|y|\mathbf{n}\cdot \boldsymbol{\sigma }\right]F\right\}.
\end{equation}
This formula corrects the exponent in (\ref{vectorPDF}). Level-crossing probability (\ref{eq:cond}) then, with the first correction in $1/\lambda ^2$
 taken into account, becomes
 \begin{equation}\label{eq:cond1}
 \frac{d\mathcal{P}_{U_N}(\lambda ,\bar{\lambda })}{d\lambda d\bar{\lambda }}
 \simeq \frac{\Delta ^3}{8N(N-1)\sqrt{\pi }\,\sigma ^3}\,\,\frac{\left|\mathop{\mathrm{Im}}\lambda \right|}{\left|\lambda \right|^6}\,\,{\rm e}\,^{-\frac{\Delta ^2}{4\sigma ^2\left|\lambda \right|^2}}
 \left\langle \,{\rm e}\,^{-\frac{1}{2\sigma ^2}\,\mathop{\mathrm{tr}}(A^{-1}F^\dagger MF)}\right\rangle_F,
\end{equation}
where $M$ is a shorthand notation for the $2\times 2$ matrix
\begin{equation}
 M:=\frac{\Delta}{x^2+y^2}\left[\left(y^2-x^2\right)\sigma _3+2x|y|\mathbf{n}\cdot \boldsymbol{\sigma }\right].
\end{equation}
The statistical average over $F$ is Gaussian with variance $\sigma^2$. The combinatorial factor $N(N-1)$ takes into account that we are concentrating on just one of the $N(N-1)$ level-crossing points.

The well-known formula for the Gaussian average of an exponential function with a quadratic exponent gives:  
\begin{equation}
 \left\langle \,{\rm e}\,^{-\frac{1}{2\sigma }\,\mathop{\mathrm{tr}}(A^{-1}F^\dagger MF)}\right\rangle_F=\det\nolimits^{-1}\left(\mathbbm{1}\otimes\mathbbm{1}+\frac{1}{2}\,A^{-1}\otimes M\right)=\prod_{k=3}^{N}\prod_{j=1,2}^{}
\left(1+\frac{m_j}{2\tilde E _k}\right)^{-1},
\end{equation}
where $m_j$ are the eigenvalues of $M$ and $\tilde E _k$ are the eigenvalues of $A$ given by \eqref{shifetspectrum}. Now, because $\mathop{\mathrm{tr}}M=0$ and $\det M=\Delta^2$ (which is easy to show using the fact that $\mathbf{n}$ is perpendicular to $\mathbf{e}_3$), the eigenvalues of $M$ are $m_j=\pm \Delta $, and
\begin{equation}\label{corfact}
 \left\langle \,{\rm e}\,^{-\frac{1}{2\sigma ^2}\,\mathop{\mathrm{tr}}(A^{-1}F^\dagger MF)}\right\rangle_F=\prod_{k=3}^{N}\left(1-\frac{\Delta ^2}{4\tilde E_k^2}\right)^{-1},
\end{equation}
which gives for the weak-coupling asymptotics of the level-crossing PDF:
 \begin{equation}\label{cond-asympt}
 \frac{d\mathcal{P}_{U_N}(\lambda ,\bar{\lambda })}{d\lambda d\bar{\lambda }}
 \simeq \frac{\Delta ^3}{8N(N-1)\sqrt{\pi }\,\sigma ^3}\,\,\frac{\left|\mathop{\mathrm{Im}}\lambda \right|}{\left|\lambda \right|^6}\,
 \prod_{k=3}^{N}\!{}^{{}}\left(1-\frac{\Delta ^2}{4\tilde E _k^2}\right)^{-1}
 \,{\rm e}\,^{-\frac{\Delta ^2}{4\sigma ^2\left|\lambda \right|^2}}.
\end{equation}
This expression is asymptotically exact in the $\lambda \rightarrow 0$ limit. As detailed above, $\Delta $ is the   distance between the closest pair of eigenvalues of the unperturbed  matrix $H_0$,  in the product these two eigenvalues are omitted, and $\tilde E_k$'s are defined in (\ref{shifetspectrum}).

\begin{figure}[t]
\begin{center}
 \subfigure[]{
   \includegraphics[height=4.1cm] {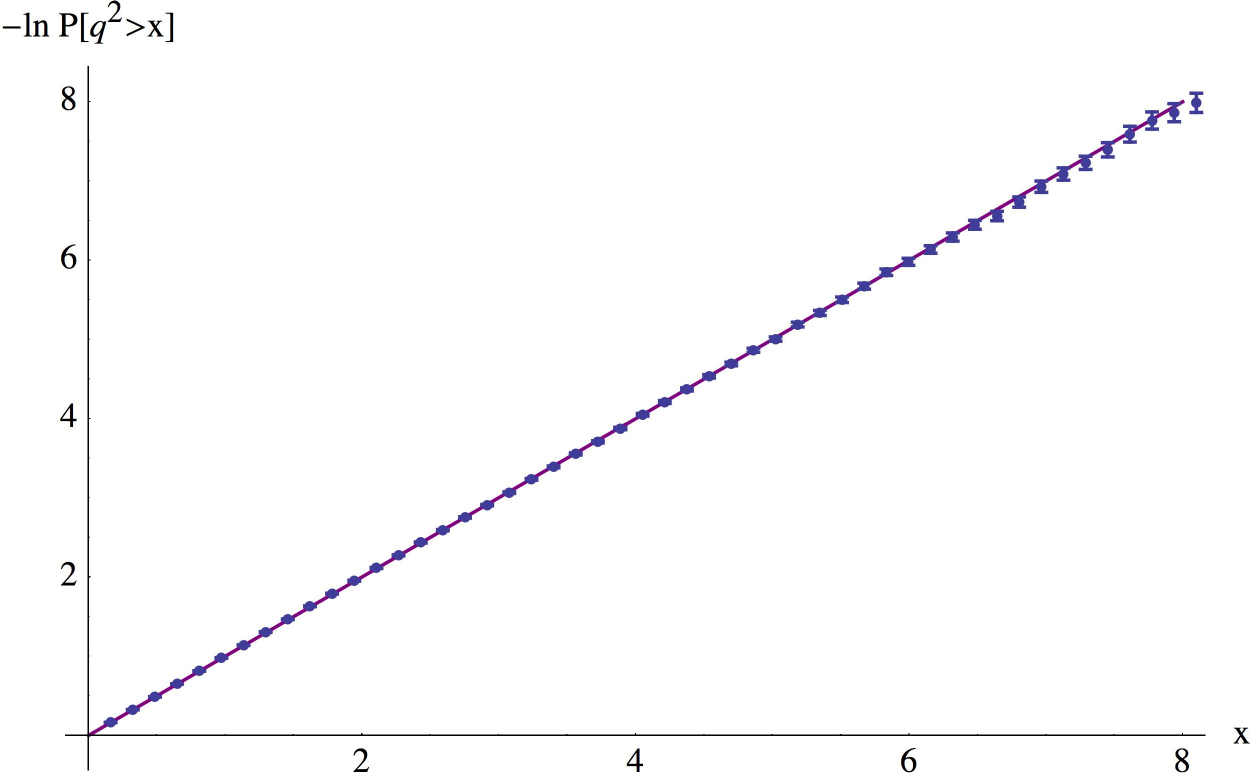}
   \label{fig2:subfig1}
 }
 \subfigure[]{
   \includegraphics[height=4.1cm] {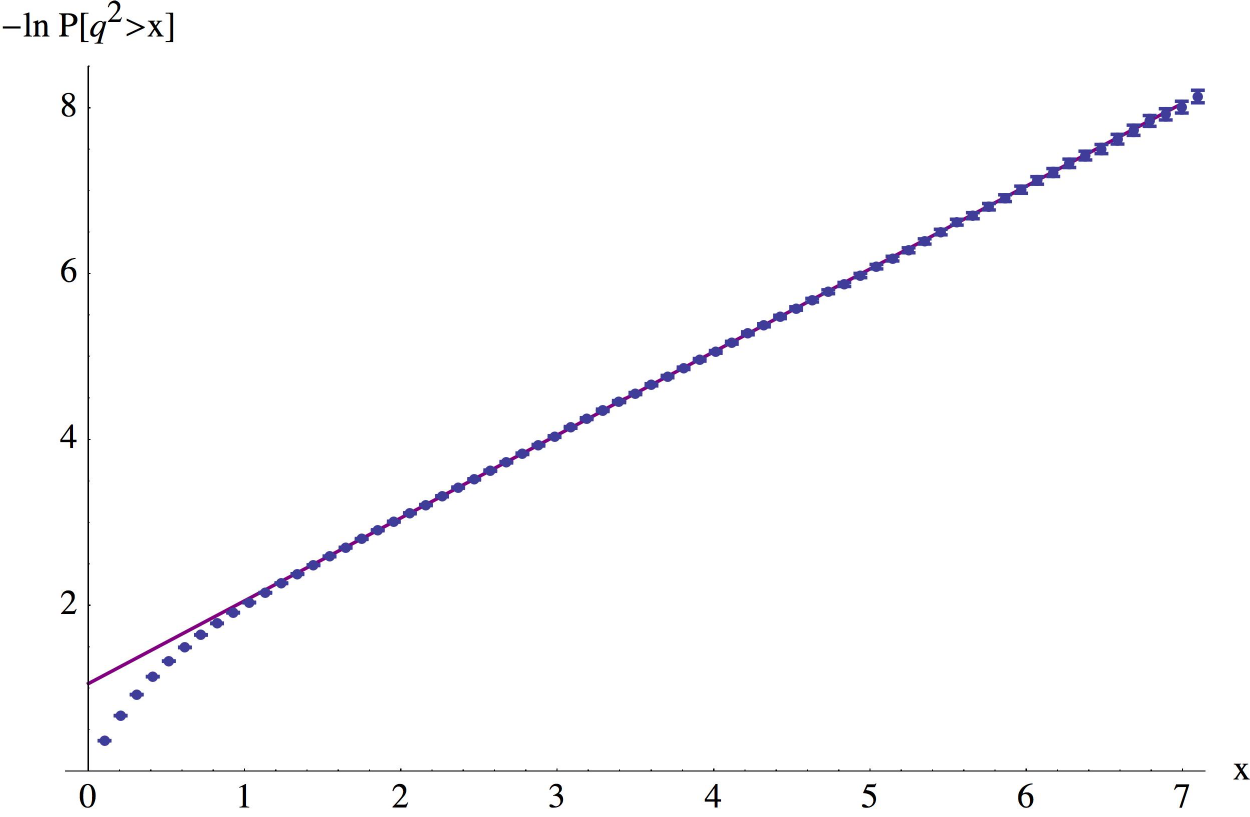}
   \label{fig2:subfig2}
 }
\caption{\label{basicfig}\small The cumulative level-crossing probability: (a) for $2\times 2$ matrices with $H_0=\mathop{\mathrm{diag}}(1,-1)$, corresponding to $\Delta =2$, and $\sigma =1$; (b) for $3\times 3$ matrices with $H_0=\mathop{\mathrm{diag}}(1,-1,5)$ and $\sigma =1$. The solid line is the asymptotic formula (\ref{asumpt}). The dots  represent numerical results.}
\end{center}
\end{figure}

The asymptotic distribution in the variable $q^2$, introduced in (\ref{p+iq}), is again given by the Poisson law but with a ``wrong" normalization constant:
\begin{equation}\label{asumpt}
 \mathcal{P}_{U_N}(q^2>x)
 \simeq \frac{2}{N(N-1)}\,
  \prod_{k=3}^N\!{}^{{}}\left(1-\frac{\Delta ^2}{4\tilde E_k^2}\right)^{-1}
 \,{\rm e}\,^{-\frac{\Delta ^2x}{4\sigma ^2}}.
\end{equation}
Observe that the right-hand side of the latter formula  does not converge to one at $x=0$. This formula is exact for $2\times 2$ matrices when it coincides with (\ref{Poisson}). 
In general, it describes the  asymptotical behavior of PDF for  large $x$, and deviates from the exact result when  $x$ is small. This is clearly visible in fig.~\ref{basicfig}, where the asymptotic formula (\ref{asumpt}) is compared to numerical data. For the $2\times 2$ matrices, fig.~\ref{fig2:subfig1}, the data perfectly agrees with the Poisson distribution in the whole range of the variable $q^2$. In the $3\times 3$ case, fig.~\ref{fig2:subfig2}, the data quickly approaches the asymptotic regime predicted by (\ref{asumpt}), but  at small $q^2$ the deviations from the Poisson distribution are clearly visible.

\begin{figure}[t]
\begin{center}
 \centerline{\includegraphics[width=8cm]{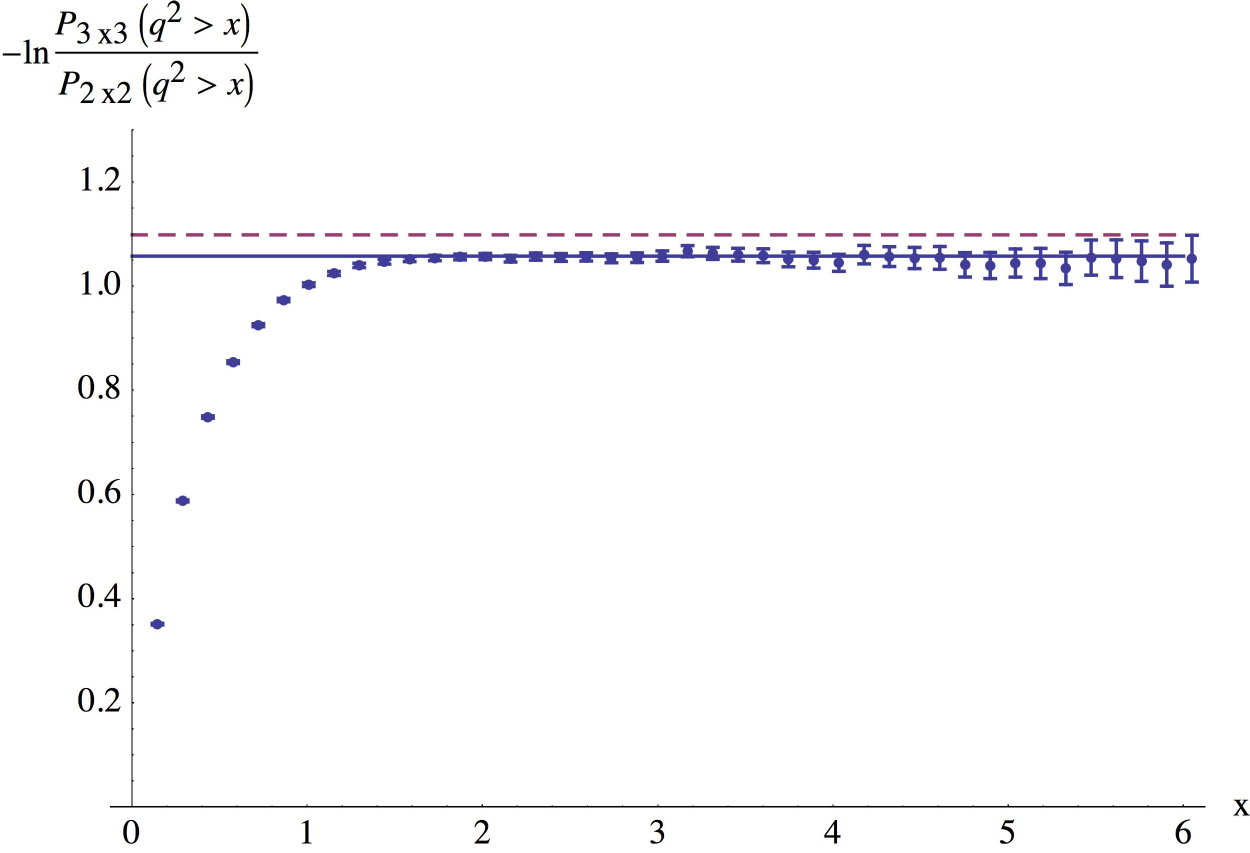}}
\caption{\label{norm-fig}\small The cumulative level-crossing probability normalized to the Poisson distribution. The values of parameters are the same as in fig.~\ref{fig2:subfig2}. The horizontal lines are minus the logarithm of the normalization factor in the asymptotic formula (\ref{asumpt}), with (the solid line) or without (the dashed line) the correction factor (\ref{corfact}). Even though the difference is very small, the numerical data clearly shows that the corrections from other eigenvalues are necessary to get the right asymptotics of the probability.}
\end{center}
\end{figure}

The correction factor due to spectator eigenvalues (\ref{corfact}) is actually very close to one. Indeed, each $|\tilde{E}_k|$ must be at least as big as $3\Delta/2 $. Otherwise the distance between $E_k$ and $E_1$ or $E_2$ would be smaller than $\Delta $ while we have assumed that $\Delta $ is the smallest gap in the spectrum. Consequently the correction factor associated with each particular eigenvalue lies between $1$ and $9/8$. The contribution of eigenvalues further away is even smaller. For instance, for parameters in fig.~\ref{fig2:subfig2}, the correction factor is $25/24$. Yet, we were able to check numerically that the correction factor is necessary to reproduce the correct asymptotics of the numerical data, as shown in fig.~\ref{norm-fig}.

\subsection{Independent collisions approximation}
\label{ICA-sec}

Calculating the level-crossing PDF exactly  is a complicated problem. It is difficult to come up with a closed expression already for $3\times 3$ matrices. Nevertheless, we have found a heuristic approximate formula which describes numerical data remarkably well in the full range of parameters. 

The idea is very simple. A collision of more than two eigenvalues happens with zero probability. Moreover, secular perturbation theory effectively reduces  level crossing to  a $2\times 2$ problem \cite{vonNeumann:1929:VEA,landau1991quantum}. The additional key assumption we make here is that  collisions of different pairs of eigenvalues are  statistically independent events. Such assumption is clearly only an approximation, not really justfied by any small parameter, but it turns out to work surprisingly well.

The total level-crossing PDF is then the sum over all pairs of eigenvalues of the partial probabilities of pairwise collisions, where each partial probability is given by eq.~(\ref{eq:cond}). We shall call this procedure the {\it Independent Collisions Approximation} (ICA). It results in a heuristic formula for matrices of any size:
\begin{equation}\label{heur}
 \frac{d\mathcal{P}_{U_N}(\lambda ,\bar{\lambda })}{d\lambda d\bar{\lambda }}
 \approx \frac{1}{8N(N-1)\sqrt{\pi }\,\sigma ^3}\,
 \,\frac{\left|\mathop{\mathrm{Im}}\lambda \right|}{\left|\lambda \right|^6}\,
 \sum_{1\le i<j\le N}^{}|E_i-E_j|^3
 \,{\rm e}\,^{-\frac{\left(E_i-E_j\right)^2}{4\sigma ^2\left|\lambda \right|^2}}.
\end{equation}

Analogously to \eqref{Poisson},  the cumulative distribution in the variable $q^2$ defined in (\ref{p+iq}) is given  by the sum of the  independent Poisson distributions for each pair of eigenvalues:
\begin{equation}\label{Poisson-anyN}
  \mathcal{P}_{U_N}(q^2>x)
 \approx \frac{2}{N(N-1)}\,
 \sum_{1\le i<j\le N}^{}
 \,{\rm e}\,^{-\frac{\left(E_i-E_j\right) ^2x}{4\sigma ^2}}.
\end{equation}

\begin{figure}[t]
\begin{center}
 \subfigure[]{
   \includegraphics[height=4cm] {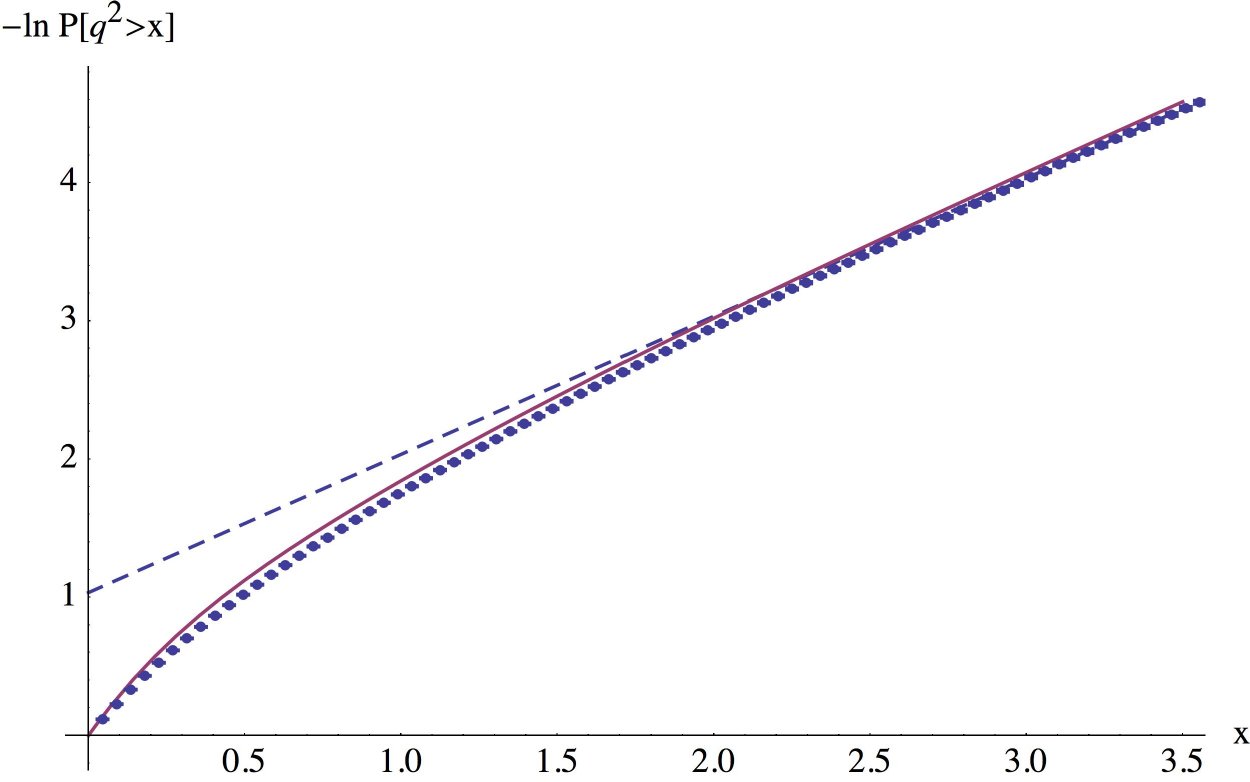}
   \label{fig2:subfig1pca}
 }
 \subfigure[]{
   \includegraphics[height=4cm] {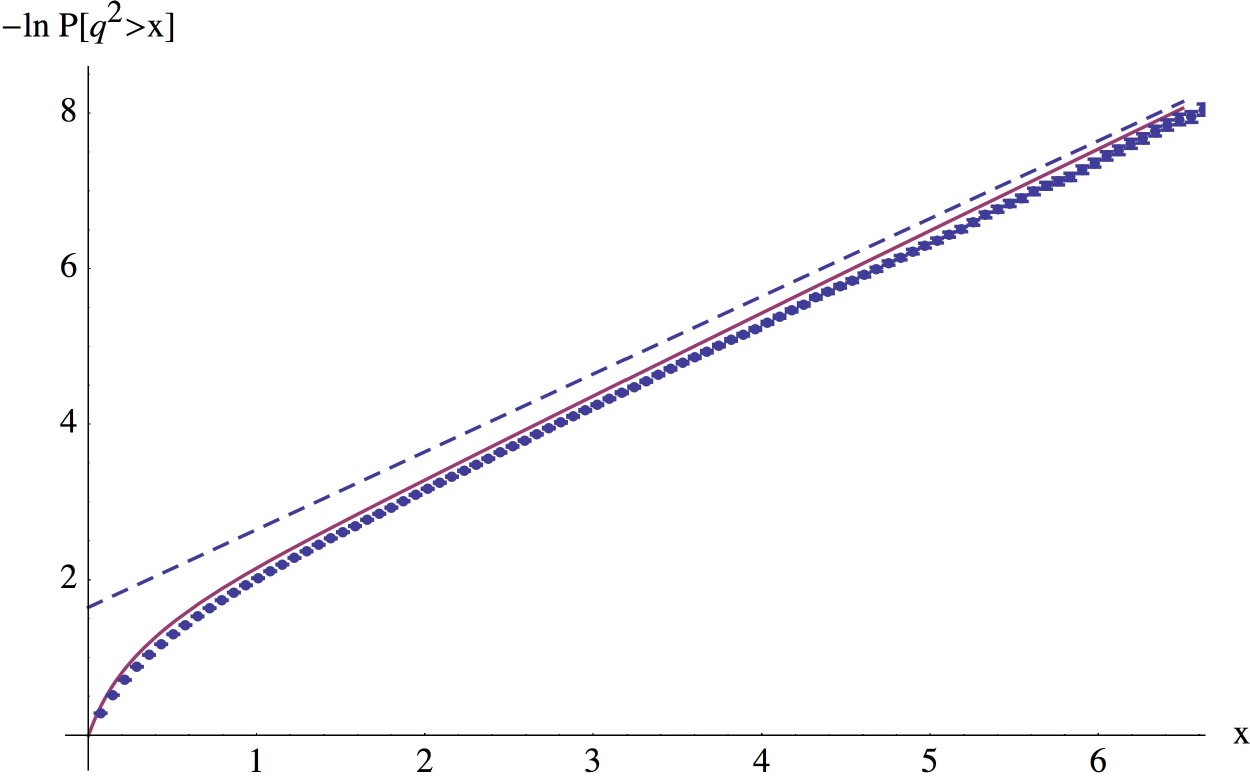}
   \label{fig2:subfig2pca}
 }
\caption{\label{PCA-fig}\small The cumulative level-crossing probability compared to the approximate ICA formula (\ref{Poisson-anyN}):
(a) for $3\times 3$ matrices with $H_0=\mathop{\mathrm{diag}}(1,-1,4)$; (b) for $4\times 4$ matrices with $H_0=\mathop{\mathrm{diag}}(1,-1,3.2,5)$. In both cases $\sigma =1$. The dashed line is the asymptotic prediction (\ref{asumpt}).}
\end{center}
\end{figure}

The formula~\eqref{Poisson-anyN} is exact for $N=2$, while for general $N$ it is only justified heuristically. Reduction to the $2\times 2$ problem is expected to give a good approximation  for small $|\lambda| $, as discussed in \S~\ref{weak-c-sec}. At the moment, we do not know of any mathematically consistent  derivation of these results for arbitrary $\lambda $ and $N$. Nevertheless, they agree with numerics reasonably well  in the whole range of $\lambda $, at  the percent level of accuracy. The comparison to numerics for $3\times 3$ and $4\times 4$ matrices is shown in fig.~\ref{PCA-fig}.

\begin{figure}[t]
\begin{center}
 \subfigure[]{
   \includegraphics[height=4cm] {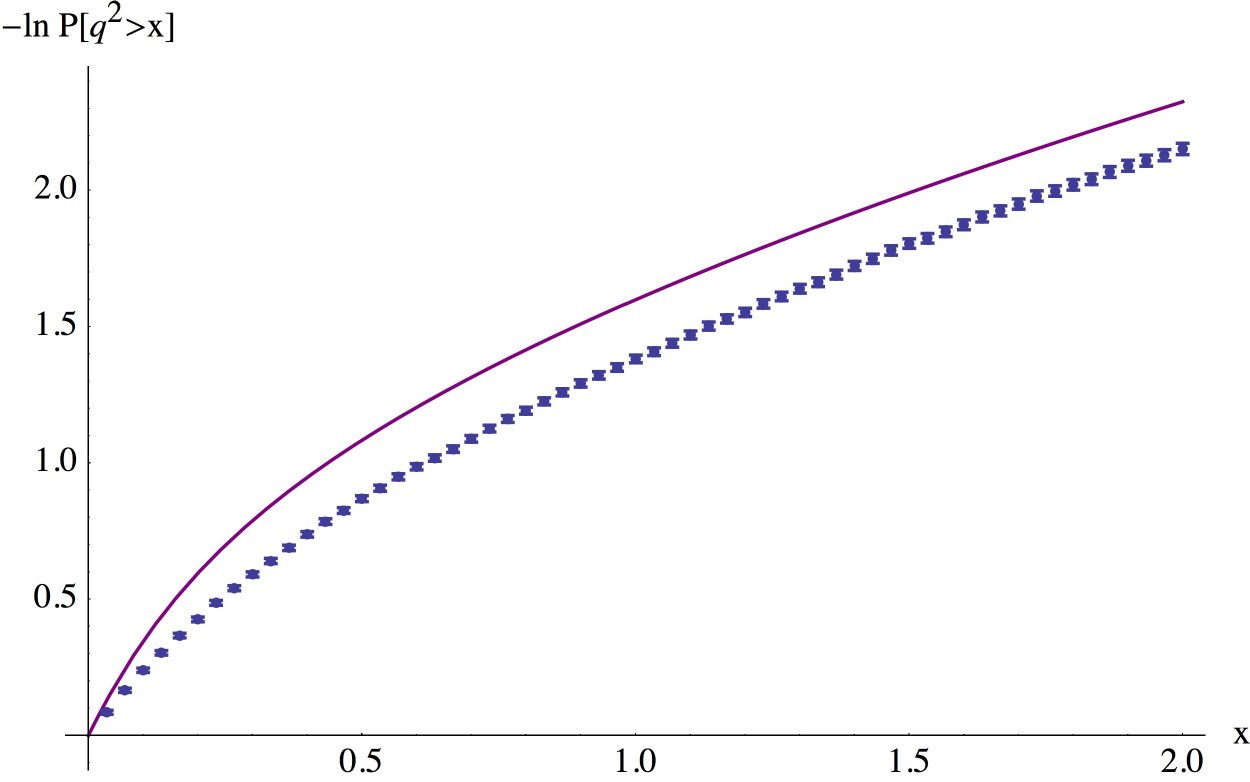}
   \label{fig2:subfig1iica}
 }
 \subfigure[]{
   \includegraphics[height=4cm] {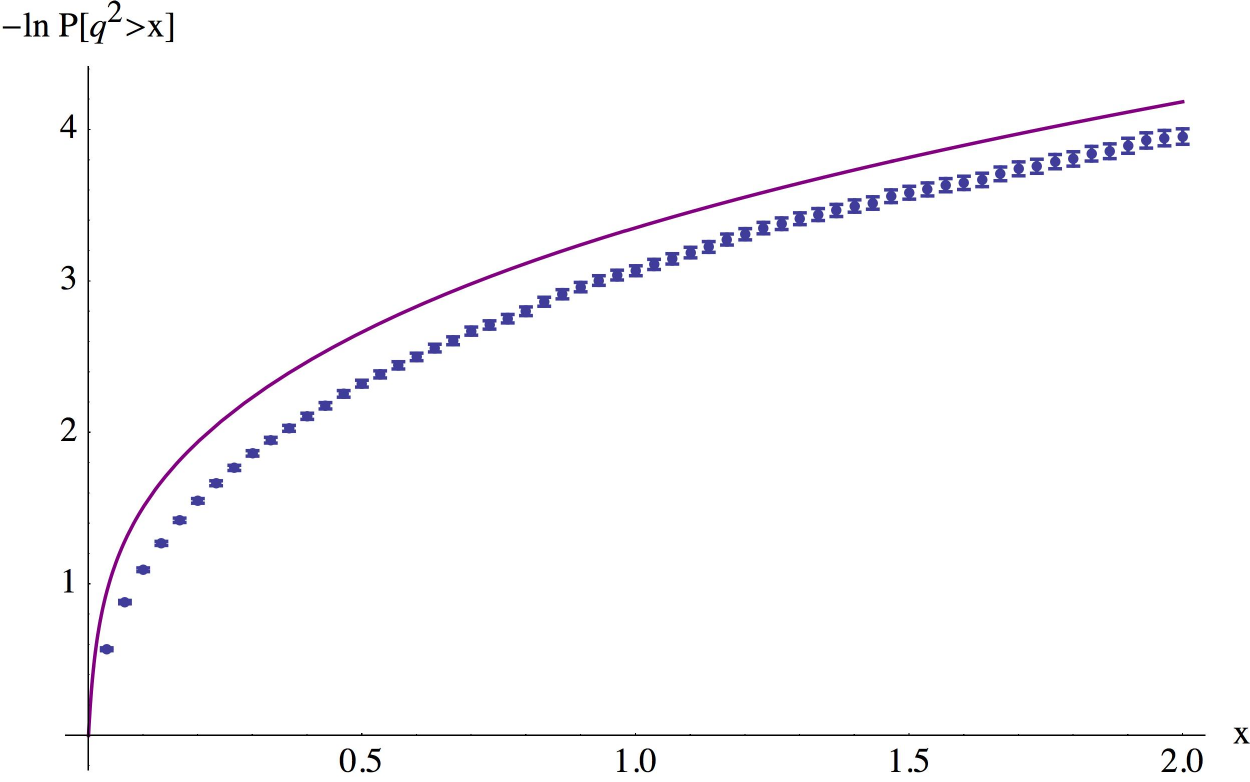}
   \label{fig2:subfig2iica}
 }
\caption{\label{ICA-fig}\small The accuracy of ICA as a function of matrix size:
(a) for $4\times 4$ matrices; (b) for $12\times 12$ matrices. The matrices are of the form $H_0=\mathop{\mathrm{diag}}(\lambda _0,\ldots ,\lambda_{N-1})$ with $\lambda _k=k+0.2k^2$, and $\sigma =1$.}
\end{center}
\end{figure}

\begin{figure}[t]
\begin{center}
 \centerline{\includegraphics[width=5cm]{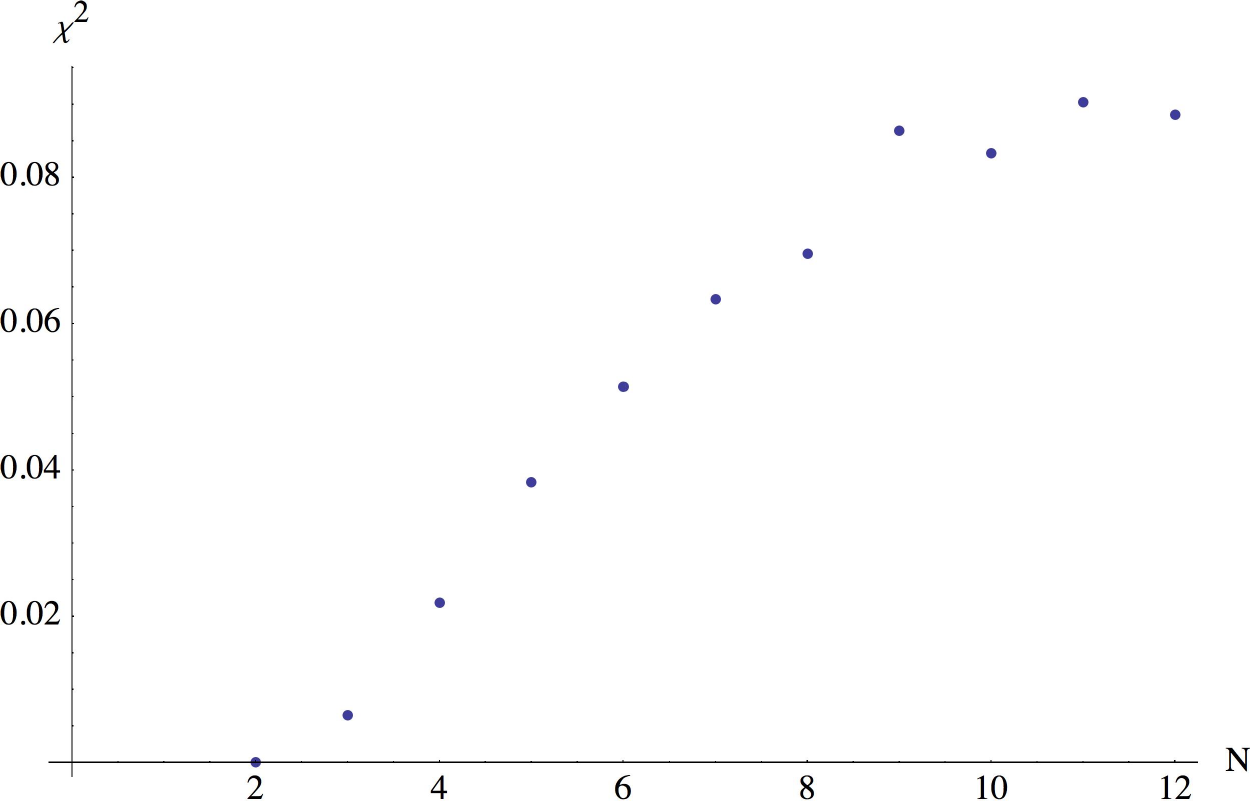}}
\caption{\label{Chi2}\small The $\chi ^2$ per point of the ICA for cumulative level-crossing probability as a function of matrix size.}
\end{center}
\end{figure}

An interesting question is how the accuracy of ICA scales with the matrix size. We have not attempted to investigate this question in full detail, but instead studied it numerically in one representative case. The results of this preliminary study are displayed in fig.~\ref{ICA-fig} and fig.~\ref{Chi2}.
We
compared ICA to numerical data for the matrix sequence of the form $H_0=\mathop{\mathrm{diag}}(\lambda _0,\ldots ,\lambda_{N-1})$, where $\lambda _k=k+0.2k^2$, up to the matrix size $N=12$. The data shows overall good agreement with  ICA: there is no much of a  difference between fig.~\ref{fig2:subfig1iica}, showing data for $N=4$, and  fig.~\ref{fig2:subfig2iica}, where the data for $N=12$ are displayed. To quantify this, in fig.~\ref{Chi2} we plot $\chi ^2$ per point for the logarithm of the cumulative probability $\ln P(q^2>x)$ for matrices of different size. The $\chi ^2$
shows a moderate growth with $N$ at small $N$, but stabilizes for $N>8$.

\section{Gaussian Orthogonal Ensemble}

Next we consider the Gaussian Orthogonal Ensemble $GOE_N$ of random real symmetric matrices: 
 \begin{equation}\label{GOE}
 d\mathcal{P}(V)=2^{-\frac{N}{2}}\left(2\pi \sigma^2 \right)^{-\frac{N(N+1)}{4}}\,{\rm e}\,^{-\frac{1}{4\sigma ^2}\,\mathop{\mathrm{tr}}V^2}\prod_{1\le i\leq j\le N}^{}dV_{ij},
\end{equation}

The PDF for $2\times 2$ matrices  was found in the pioneering paper \cite{zirnbauer1983destruction} and is given by: 
\begin{equation}\label{eq:cond2}
 d\mathcal{P}_{O_2}(\lambda ,\bar{\lambda })
 =\frac{\Delta ^2}{16\pi \sigma ^2}\,\,\frac{1}{\left|\lambda \right|^4}\,\,{\rm e}\,^{-\frac{\Delta ^2}{8\sigma ^2\left|\lambda \right|^2}}d\lambda d\bar{\lambda }.
\end{equation}
Interestingly,  PDF is rotation-invariant, i.e. depends on $|\lambda|$ but not on the argument of $\lambda$.  Let us perhaps mention how this important difference to $GUE_2$ comes about. The level-crossing condition is the same equation (\ref{v+a+ib}), with $\mathbf{a}$ and $\mathbf{b}$ given by the first terms in eq.~(\ref{ab-to-lineraorder}). However, symmetric traceless $2\times 2$ matrices expand in the two-dimensional basis $\{\sigma _1,\sigma _3\}$ and all the vectors involved belong to $\mathbbm{R}^2$. The solution (\ref{v=bn-a}), (\ref{bn-eq}) is a set of two points, and the prefactor in the level-crossing PDF is just the Jacobian of transformation from $\mathbf{v}$ to $\lambda ,\bar{\lambda }$, proportional to $1/|\lambda |^4$.

In the variables $(p,q)$ from \eqref{p+iq}, we get: 
\begin{equation}\label{eq:cond2pq}
 d\mathcal{P}_{O_2}(p,q)
 =\frac{\Delta ^2}{8\pi \sigma ^2}\,\,{\rm e}\,^{-\frac{\Delta ^2(p^2+q^2)}{8\sigma ^2}}dp dq.
\end{equation}
Now it is the radial distribution that is given by the Poisson density. The cumulative radial probability is 
\begin{equation}\label{Poisson-GOE}
  \mathcal{P}_{O_2}(p^2+q^2>x)
 =\,{\rm e}\,^{-\frac{\Delta ^2x}{8 \sigma ^2}}.
\end{equation}

\begin{figure}[t]
\begin{center}
 \centerline{\includegraphics[width=8cm]{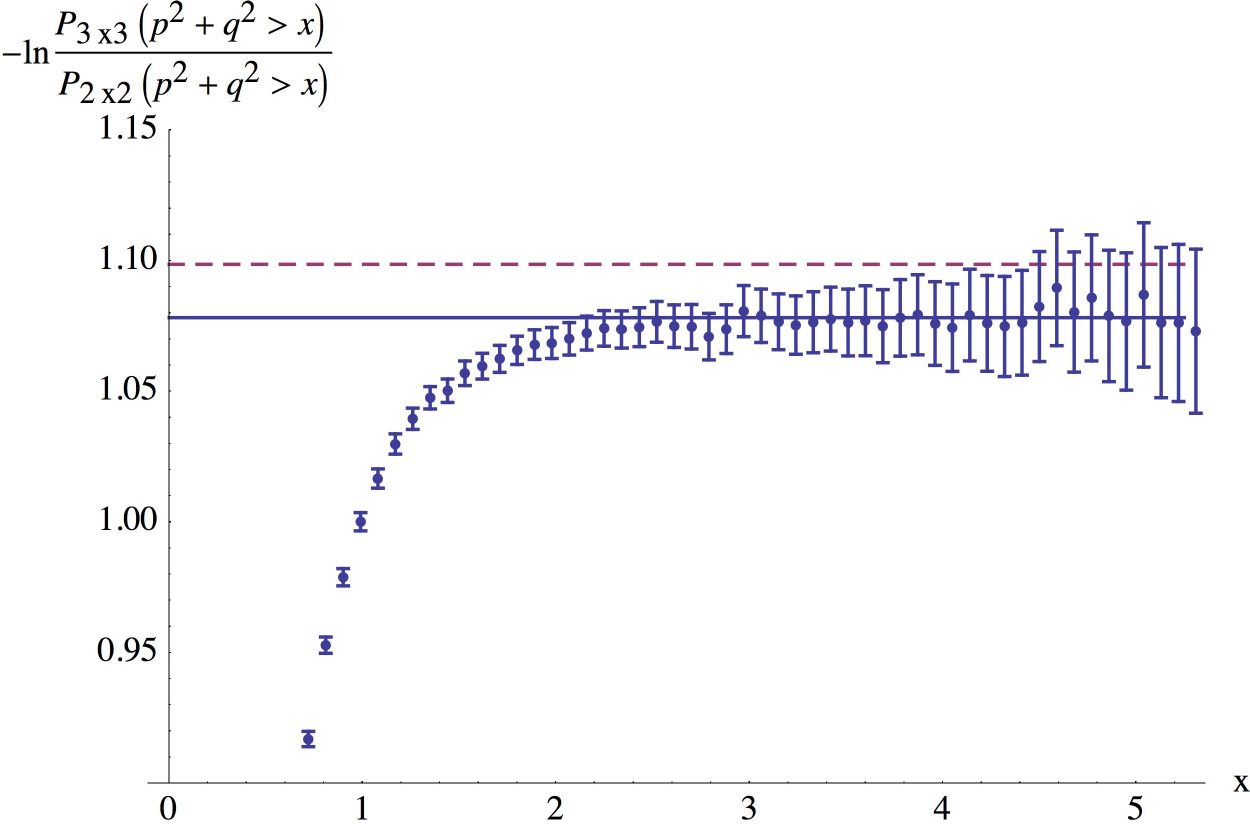}}
\caption{\label{Norm-GOE}\small Cumulative radial probability normalized to the Poisson distribution for $GOE$ with $H_0=\mathop{\mathrm{diag}}(1,-1,5)$ and $\sigma =1/\sqrt{2}$. The blue straight line is the asymptotic prediction (\ref{eq:cond-asympt-GOE}). The dashed line is the would-be asymptotics without the correction factor (\ref{corfact-GOE}).}
\end{center}
\end{figure}

The small-$\lambda $  asymptotics of PDF for any $N$ is governed by the smallest gap in the spectrum. The derivation is the same as in sec.~\ref{weak-c-sec}, except that now the rectangular matrix $F$ is real and
\begin{equation}\label{corfact-GOE}
 \left\langle \,{\rm e}\,^{-\frac{1}{2\sigma ^2}\,\mathop{\mathrm{tr}}(A^{-1}F^t MF)}\right\rangle_F=\prod_{k=3}^{N}\left(1-\frac{\Delta ^2}{4\tilde E_k^2}\right)^{-\frac{1}{2}}.
\end{equation}
The result is
\begin{equation}\label{eq:cond-asympt-GOE}
 d\mathcal{P}_{O_N}(\lambda ,\bar{\lambda })
 \simeq\frac{\Delta ^2}{8\pi N(N-1)\sigma ^2}\,\,\frac{1}{\left|\lambda \right|^4}
 \,
  \prod_{k=3}^N
 \left(1-\frac{\Delta^2}{4\tilde E_k^2} \right)^{-\frac{1}{2}}\,{\rm e}\,^{-\frac{\Delta^2}{8\sigma ^2\left|\lambda \right|^2}}d\lambda d\bar{\lambda }.
\end{equation}
We confirmed numerically that the asymptotic PDF has correct normalization. In fig.~\ref{Norm-GOE} the cumulative radial probability normalized to the Poisson distribution is plotted against the asymptotic prediction.

\begin{figure}[t]
\begin{center}
 \subfigure[]{
   \includegraphics[height=3.95cm] {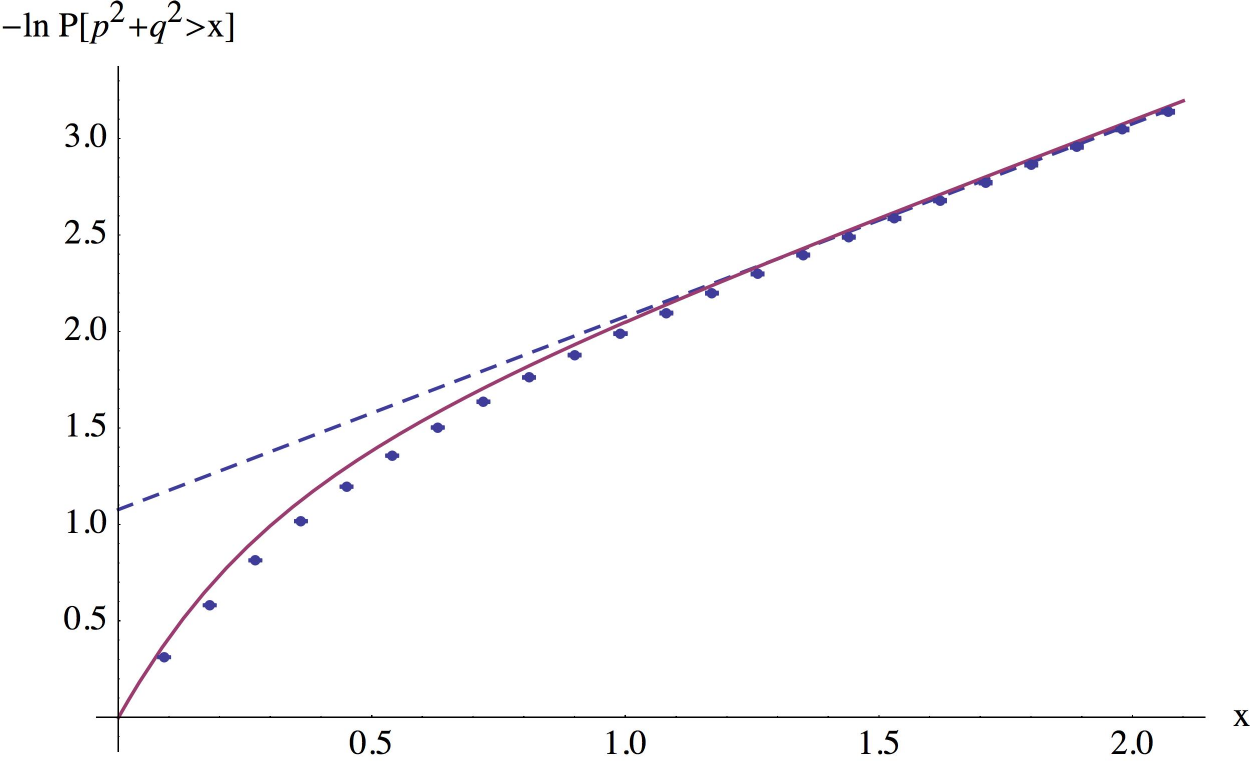}
   \label{fig2:subfig1pca}
 }
 \subfigure[]{
   \includegraphics[height=3.95cm] {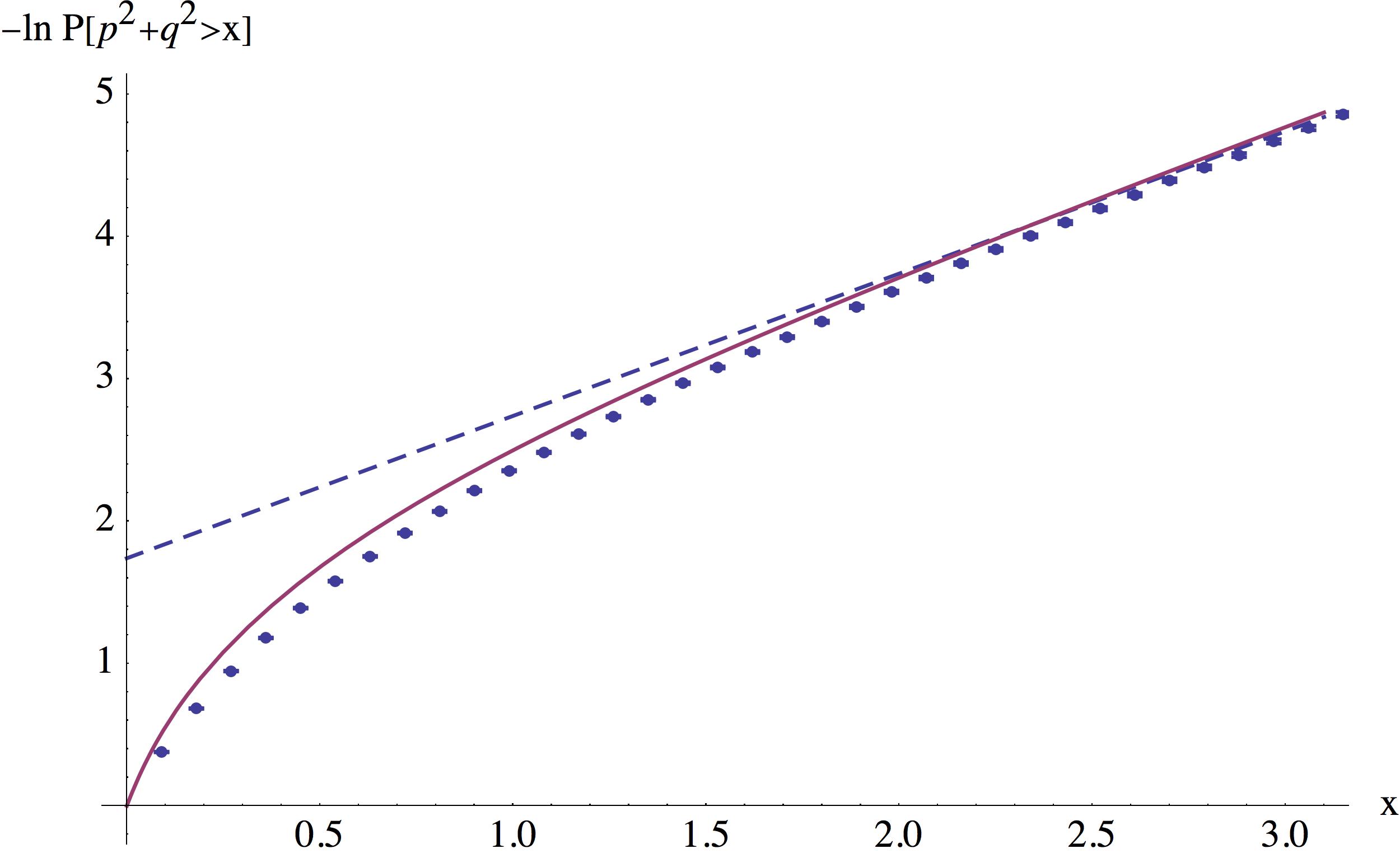}
   \label{fig2:subfig2pca}
 }
\caption{\label{PCA-fig}\small The cumulative radial probability for GOE compared to ICA:
(a) for $3\times 3$ matrices and the same parameters as in  fig.~\ref{Norm-GOE}:  $H_0=\mathop{\mathrm{diag}}(1,-1,5)$,  $\sigma =1/\sqrt{2}$; (b) for $4\times 4$ matrices with $H_0=\mathop{\mathrm{diag}}(1,-1,5,-4)$ and  $\sigma =1/\sqrt{2}$.  The dashed line is the Poisson asymptotics.}
\end{center}
\end{figure}

The whole PDF, at any $N$ and for any $\lambda $, can be described, with reasonable accuracy, by ICA discussed in sec.~\ref{ICA-sec}. The PDF in this approximation is given by an additive combination of the partial two-level probabilities:
\begin{equation}\label{eq:cond3}
 d\mathcal{P}_{O_N}(\lambda ,\bar{\lambda })
 \simeq\frac{1}{8\pi N(N-1)\sigma ^2}\,\,\frac{1}{\left|\lambda \right|^4}
 \,
  \sum_{1\le i<j\le N}^{}
 \left(E_i-E_j\right)^2\,{\rm e}\,^{-\frac{\left(E_i-E_j\right) ^2}{8\sigma ^2\left|\lambda \right|^2}}d\lambda d\bar{\lambda }.
\end{equation}
We checked numerically that this is indeed a good approximation. In fig.~\ref{GOE-ICA} the cumulative radial probability constructed from ICA is compared to numerical data for $3\times 3$ matrices.

\begin{figure}[t]
\begin{center}
   \includegraphics[width=8cm] {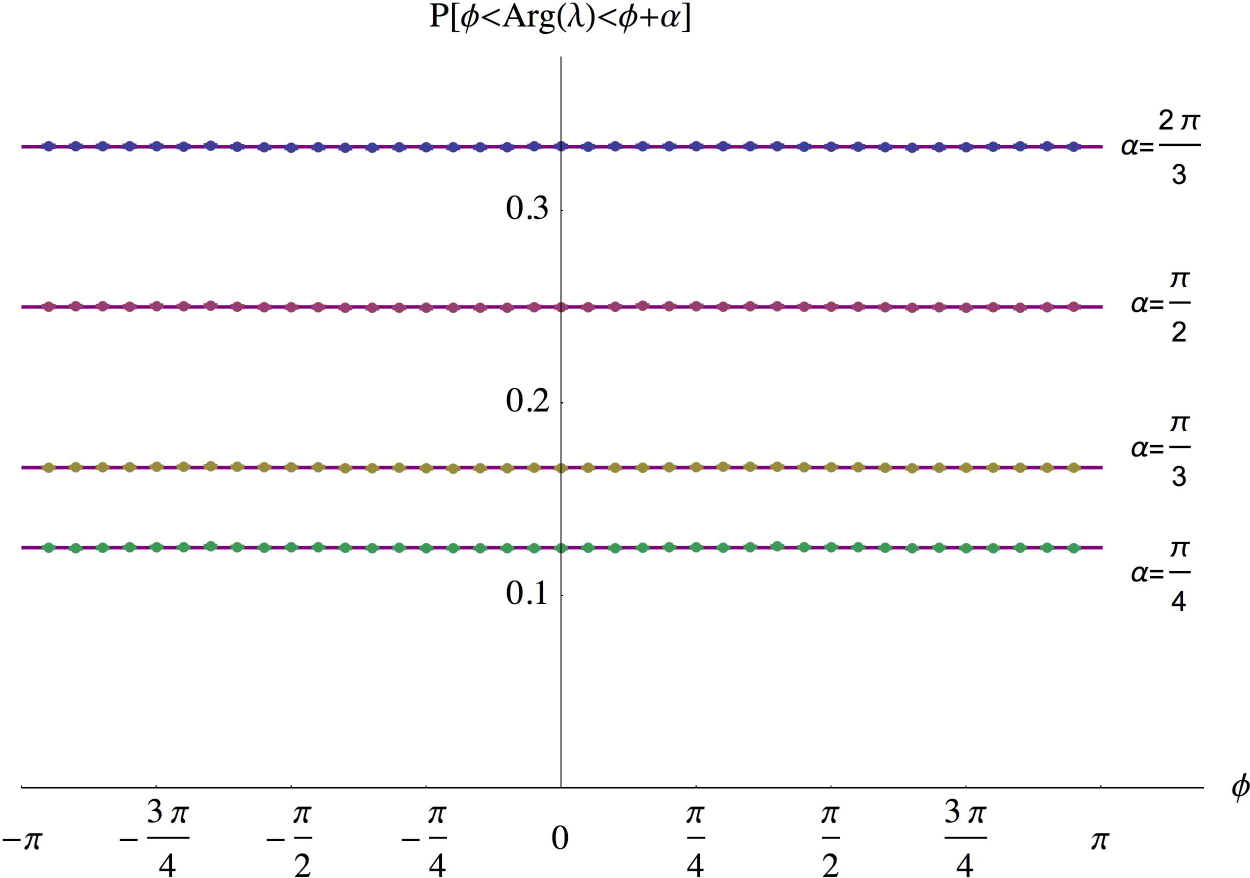}
\caption{\label{GOE3x3-angular}\small  The angular dependence of the probability distributions
 for $3\times 3$ matrices.  The parameters are the same as in fig.~\ref{Norm-GOE}.}
\end{center}
\end{figure}

Explicit calculation shows that the level-crossing probability for $2\times 2$ matrices from $GOE$ depends only on the absolute value of $\lambda $. ICA inherits this property, but this is only an approximation. An interesting question is whether the exact level-crossing PDF for $N>2$ is rotationally invariant.
We have tested the rotational symmetry of the PDF by numerically calculating the probability of the level crossing to lie in the sector $\phi <\mathop{\mathrm{Arg}}\lambda <\phi +\alpha $. For a rotationally symmetric PDF this probaility does not depend on $\phi $:
\begin{equation}
 \mathcal{P}\left(\phi <\mathop{\mathrm{Arg}}\lambda <\phi +\alpha\right)
 \stackrel{{\rm rot.symm.}}{\Longrightarrow}\frac{\alpha }{2\pi }\,.
\end{equation}
The numerical results for $3\times 3$ matrices, shown in fig.~\ref{GOE3x3-angular} perfectly agree with this assumption. This agreement cannot be attributed to the accidental accuracy of ICA, which is rotationally symmetric by construction. Figs.~\ref{GOE-ICA} and \ref{GOE3x3-angular} represent the same data, and while the deviations from ICA in fig.~\ref{GOE-ICA} are small they are clearly visible. At the same time the angular probability in fig.~\ref{GOE3x3-angular} is perfectly flat, with deviations smaller than errorbars.

We are led to conclusion, which we put forward as a conjecture, that the distribution of level-crossing points in $GOE_N$ is invariant under rotations
\begin{equation}
 \lambda \rightarrow \,{\rm e}\,^{i\varphi }\lambda .
\end{equation}
For $N=2$ this follows from the explicit calculation, but so far we could not find any appropriate symmetry which might explain this phenomenon. 

\section{General complex and real  matrices}

The other two random matrix ensembles that we consider are the general complex Gaussian matrices $GE^\bC_N$, with the probability density:
\begin{equation}\label{Gauss2}
 d\mathcal{P}(V)=\left(8\pi \sigma^2 \right)^{-N^2}\,{\rm e}\,^{-\frac{1}{4\sigma ^2}\,\mathop{\mathrm{tr}}V^\dagger V}\prod_{i j}^{}dV_{ij}dV^*_{ij},
\end{equation}
and general real Gaussian matrices $GE^\bR_N$ with the density:
\begin{equation}\label{Gauss3}
 d\mathcal{P}(V)=\left(4\pi \sigma^2 \right)^{-\frac{N^2}{2}}\,{\rm e}\,^{-\frac{1}{4\sigma ^2}\,\mathop{\mathrm{tr}}V^tV}\prod_{i j}^{}dV_{ij},
\end{equation}
For these two ensembles we will restrict ourselves to the derivation of the level-crossing PDF for the $2\times 2$ case. 

\subsection {Complex matrices} 

The expansion coefficients in the Pauli matrices (\ref{Pauliexp}) are now complex vectors. The level crossing occurs under the condition that the complex vector $\mathbf{v}+\Delta \mathbf{e}_3/2\lambda $ is null. We can thus write the level-crossing PDF as
\begin{eqnarray}
 d\mathcal{P}_{\mathbbm{C}^2}(\lambda ,\bar{\lambda })&=&\frac{\Delta ^2}{2}\,\,
 \frac{d\lambda d\bar{\lambda }}{|\lambda |^4}\,
 \left\langle 
 \left(v_3+\frac{\Delta }{2\lambda }\right)
 \left(\bar{v}_3+\frac{\Delta }{2\bar{\lambda }}\right)
 \vphantom{\delta \left(\left(\mathbf{v}+\frac{\Delta }{2\lambda }\,\mathbf{e}_3\right)^2\right)}
 \right.
\nonumber \\
&&\times \left.
 \delta \left(\left(\mathbf{v}+\frac{\Delta }{2\lambda }\,\mathbf{e}_3\right)^2\right)
 \delta \left(\left(\bar{\mathbf{v}}+\frac{\Delta }{2\bar{\lambda }}\,\mathbf{e}_3\right)^2\right)
 \right\rangle,
\end{eqnarray}
where $\left\langle \ldots\right\rangle$ denotes Gaussian average in $\mathbf{v},\bar{\mathbf{v}}$. Shifting the integration variables $\mathbf{v}\rightarrow \mathbf{v}-\Delta \mathbf{e}_3/2\lambda$, $\bar{\mathbf{v}}\rightarrow \bar{\mathbf{v}}-\Delta \mathbf{e}_3/2\bar{\lambda}$, we get:
\begin{equation}
 d\mathcal{P}_{\mathbbm{C}^2}(\lambda ,\bar{\lambda })=\frac{\Delta ^2}{32\pi ^3\sigma ^6}\,\,
 \frac{d\lambda d\bar{\lambda }}{|\lambda |^4}\,
 \int
 d^3v\,d^3\bar{v}\,v_3\bar{v}_3\delta \left(\mathbf{v}^2\right)
 \delta \left(\bar{\mathbf{v}}^2\right)
 \,{\rm e}\,^{
 -\frac{1}{2\sigma ^2}\left(\bar{\mathbf{v}}-\frac{\Delta }{2\bar{\lambda }}\,\mathbf{e}_3\right)
 \left(\mathbf{v}-\frac{\Delta }{2\lambda }\,\mathbf{e}_3\right)
 }.
\end{equation}
In the parameterization $\mathbf{v}=\,{\rm e}\,^{-i\phi }(\mathbf{r}+i\mathbf{s})$, where $\mathbf{r}$ and $\mathbf{s}$ are real vectors, and $\phi $ is the argument of $\lambda $, the last expression becomes
\begin{eqnarray}
 d\mathcal{P}_{\mathbbm{C}^2}(\lambda ,\bar{\lambda })&=&\frac{\Delta ^2}{64\pi ^3\sigma ^6}\,\,
 \frac{d\lambda d\bar{\lambda }}{|\lambda |^4}\,
 \,{\rm e}\,^{-\frac{\Delta ^2}{8\sigma ^2|\lambda |^2}}
 \int
d^3r\,d^3s\,
\left(r_3^2+s_3^2\right)\delta \left(\mathbf{r}^2-\mathbf{s}^2\right)
\delta \left(\mathbf{r}\cdot \mathbf{s}\right)
\nonumber \\
&&
 \vphantom{\frac{\Delta ^2}{64\pi ^3\sigma ^6}\,\,
 \frac{d\lambda d\bar{\lambda }}{|\lambda |^4}\,
 \,{\rm e}\,^{-\frac{\Delta ^2}{8\sigma ^2|\lambda |^2}}
 \int
d^3r\,d^3s\,
\left(r_3^2+s_3^2\right)\delta \left(\mathbf{r}^2-\mathbf{s}^2\right)
\delta \left(\mathbf{r}\cdot \mathbf{s}\right)}
\times 
\,{\rm e}\,^{-\frac{\mathbf{r}^2+\mathbf{s}^2}{2\sigma ^2}+\frac{\Delta r_3}{2\sigma ^2|\lambda |}}.
\end{eqnarray}
Let us first integrate over $\mathbf{s}$ using the two constraints in the delta functions. The constraints are solved by $\mathbf{s}=r\mathbf{n}$, where $\mathbf{n}$ is a unit vector perpendicular to $\mathbf{r}$. The solution forms a circle in the plane perpendicular to $\mathbf{r}$, which can be parameterized by the angle $\varphi $. In particular, $s_3=\sqrt{r^2-r_3^2}\,\cos\varphi  $. So,
\begin{eqnarray}
  d\mathcal{P}_{\mathbbm{C}^2}(\lambda ,\bar{\lambda })&=&\frac{\Delta ^2}{128\pi ^3\sigma ^6}\,\,
 \frac{d\lambda d\bar{\lambda }}{|\lambda |^4}\,
 \,{\rm e}\,^{-\frac{\Delta ^2}{8\sigma ^2|\lambda |^2}}
 \int
\frac{d^3r}{r}\,\,
\,{\rm e}\,^{-\frac{r^2}{\sigma ^2}+\frac{\Delta r_3}{2\sigma ^2|\lambda |}}
\nonumber \\
&&\times 
\int_{0}^{2\pi }d\varphi \,\left(r^2\cos^2\varphi+r_3^2\sin^2\varphi  \right)
\nonumber \\
&=&
 \frac{\Delta ^2}{128\pi ^2\sigma ^6}\,\,
 \frac{d\lambda d\bar{\lambda }}{|\lambda |^4}\,
 \,{\rm e}\,^{-\frac{\Delta ^2}{8\sigma ^2|\lambda |^2}}
 \int
\frac{d^3r}{r}\,\,\left(r^2+r_3^2\right)
\,{\rm e}\,^{-\frac{r^2}{\sigma ^2}+\frac{\Delta r_3}{2\sigma ^2|\lambda |}}
\end{eqnarray}

The remaining integral over $\mathbf{r}$ can be calculated in the spherical coordinates, and finally we obtain:
\begin{equation}
  d\mathcal{P}_{\mathbbm{C}^2}(\lambda ,\bar{\lambda })=\frac{\Delta ^2}{32\pi\sigma ^2}\,\,
 \frac{d\lambda d\bar{\lambda }}{|\lambda |^4}\,
 \Phi \left(\frac{\Delta ^2}{16\sigma ^2|\lambda |^2}\right),
\end{equation}
where
\begin{equation}
 \Phi (u)=\frac{\sqrt{\pi }}{8u^{\frac{3}{2}}}\left(4u^2+1\right)
 \,{\rm e}\,^{-u}\mathop{\mathrm{erf}}\sqrt{u}+\frac{2u-1}{u}\,\,{\rm e}\,^{-2u}.
\end{equation}
The level-crossing PDF is rotationally symmetric and depends only on $|\lambda |$, which in this case follows from the invariance of the $GE^{\mathbbm{C}}_N$ probability measure under phase transformations: $V\rightarrow \,{\rm e}\,^{-i\phi }V$.

\begin{figure}[t]
\begin{center}
   \includegraphics[width=8cm] {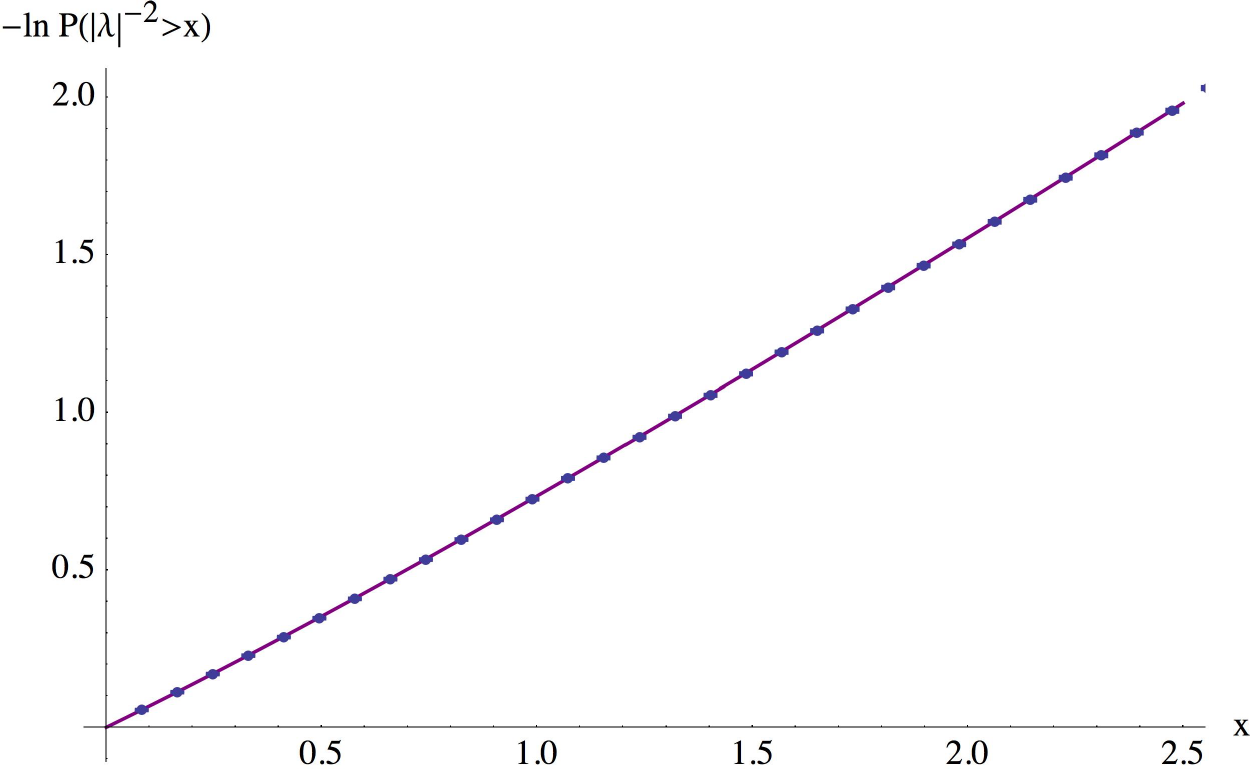}
\caption{\label{Clx}\small  The radial probability for complex $2\times 2$ matrices (\ref{radcom}), with $H_0=\mathop{\mathrm{diag}}(1,-1)$ and $\sigma =1/2$. Blue dots are numerical data, shown for comparison.}
\end{center}
\end{figure}

The cumulative radial probability is given  by
\begin{equation}\label{radcom}
 \mathcal{P}_{\mathbbm{C}_2}\left(\frac{1}{|\lambda |^2}>\frac{16\sigma^2x}{\Delta ^2}\right)=\int_{x}^{\infty }du\,\Phi (u)
 =\frac{1}{4}\,\sqrt{\frac{\pi }{x}}\left(2x+1\right)\,{\rm e}\,^{-x}\mathop{\mathrm{erf}}\sqrt{x}+\frac{1}{2}\,\,{\rm e}\,^{-2x},
\end{equation}
which is plotted in fig.~\ref{Clx}.

\subsection{Real matrices}
\label{realms}

The case of general real matrices is qualitatively different from complex or Hermitian matrices considered above, because perturbative $V$ is not Hermintian any more and there is no obstacle for the eigenvalues to cross even if $\lambda $ is real   \cite{BEDPSW}.

As before we expand the random matrix $V$ in the basis of Pauli matrices, but to make the expansion coefficients real we now multiply the imaginary Pauli matrix  $\sigma _2$ by $i$:
\begin{equation}
 V=v_{\mathbbm{1}}\mathbbm{1}+v_1\sigma _1+iv_2\sigma _2+v_3\sigma _3.
\end{equation}
The coefficients $v_i$ again form a three-dimension Gaussian random vector with variance $\sigma$,  but the level-crossing condition now changes because of the imaginary $i$ in front of the $\sigma _2$. The Euclidean scalar product in (\ref{h2=0}) transforms into the Lorentzian one. In this and the next subsections the dot-product will therefore refer to the Lorentzian quadratic form:
\begin{equation}\label{Lorentz}
 \mathbf{a}\cdot \mathbf{b}=a_1b_1-a_2b_2+a_3b_3,\qquad \mathbf{a}^2=\mathbf{a}\cdot \mathbf{a}.
\end{equation}
The level-crossing condition is then
\begin{equation}\label{vgrill}
 \left(\mathbf{v}+\frac{\Delta }{2\lambda }\,\mathbf{e}_3\right)^2=0,
\end{equation}
and the  level-crossing PDF is given by 
\begin{eqnarray}
 d\mathcal{P}_{\mathbbm{R}^2}(\lambda ,\bar{\lambda })&=&
\frac{\Delta ^2}{2}\,\,
 \frac{d\lambda d\bar{\lambda }}{|\lambda |^4}\,
\left\langle 
 \left(v_3+\frac{\Delta }{2\lambda }\right)
  \left(v_3+\frac{\Delta }{2\bar{\lambda} }\right)
      \right.
\nonumber \\
 &&\times \left.
  \delta \left(\left(\mathbf{v}+\frac{\Delta }{2\lambda }\,\mathbf{e}_3\right)^2\right)
  \delta \left( \left(\mathbf{v}+\frac{\Delta }{2\bar{\lambda} }\,\mathbf{e}_3\right)^2\right)
\right\rangle.
\end{eqnarray}
The probability measure at the same time depends on the Euclidean norm of $\mathbf{v}$, and the Euclidean scalar product will also show up in the intermediate calculations. The Euclidean product of two vectors $\mathbf{u}$ and $\mathbf{v}$ will be denoted by  $(\mathbf{u},\mathbf{v})$.

The solution to (\ref{vgrill}) has two branches. One is a curve:
\begin{equation}
 v_3=-\frac{\Delta x}{2|\lambda |^2}\,,\qquad 
 v_1^2-v_2^2=\frac{\Delta ^2y^2}{4|\lambda |^4}\,,
\end{equation}
where, as before, $x+iy=\lambda $. The other solution exists only when $\lambda $ is real and forms a two-dimensional surface
\begin{equation}
 y=0,\qquad v_1^2-v_2^2+\left(v_3+\frac{\Delta }{2\lambda}\right)^2=0.
\end{equation}
When $V$ is a general real matrix,
 $H$ is not Hermitian for real $\lambda $ and levels no longer repel. And indeed, for each realization of the random matrix $V$, the two level-crossing points either form a complex conjugate pair, or both lie on the real axis. These two possibilities are realized with equal probability. 

The probability density, consequently, has two strata:
\begin{eqnarray}
 \frac{d\mathcal{P}_{\mathbbm{R}^2}}{d\lambda d\bar{\lambda }}&=&
  \frac{\Delta ^3|y|}{16|\lambda |^6}\,
 \left\langle 
 \delta \left(v_3+\frac{\Delta x}{2|\lambda |^2}\right)
 \delta \left(v_1^2-v_2^2-\frac{\Delta ^2y^2}{4|\lambda |^4}\right)
 \right\rangle
 \nonumber \\
&&+
 \frac{\Delta }{4\lambda ^2}\,
 \delta \left(y\right)
 \left\langle 
 \left|v_3+\frac{\Delta }{2\lambda}\right|
 \delta \left(v_2^2-v_1^2-\left(v_3+\frac{\Delta }{2\lambda}\right)^2\right)
 \right\rangle.
\end{eqnarray}
The expectation values here can be computed with the help of the following formula for the Gaussian average over $v_1$, $v_2$:
\begin{equation}
 \left\langle \delta \left(v_1^2-v_2^2-u^2\right)\right\rangle_{v_1,v_2}
 =\frac{\,{\rm e}\,^{-\frac{u^2}{2\sigma ^2}}}{2\pi \sigma ^2}
 \int_{-\infty }^{+\infty }\frac{dv_2}{\sqrt{v^2_2+u^2}}\,\,
 \,{\rm e}\,^{-\frac{v^2_2}{\sigma ^2}}=\frac{K_0\left(\frac{u^2}{2\sigma ^2}\right)}{2\pi \sigma ^2}\,,
\end{equation}
where $K_\nu $ is the modified Bessel function of the second kind. We thus have
\begin{eqnarray}
 \frac{d\mathcal{P}_{\mathbbm{R}^2}}{d\lambda d\bar{\lambda }}&=&
 \frac{1}{\left(2\pi \sigma^2\right)^{\frac{3}{2}} }\int_{-\infty }^{+\infty }dv\,\,{\rm e}\,^{-\frac{v^2}{2\sigma ^2}}
 \left[
 \frac{\Delta ^3|y|}{16|\lambda |^6}\,K_0\left(\frac{\Delta ^2y^2}{8\sigma ^2|\lambda |^4}\right)\delta \left(v+\frac{\Delta x}{2|\lambda |^2}\right)
 \right.
\nonumber \\
 &&+\left.
 \frac{\Delta }{4\lambda^2}\,
 \delta \left(y\right)
 \left|v+\frac{\Delta }{2\lambda}\right|
 K_0\left(\frac{1}{2\sigma ^2}\,\left(v+\frac{\Delta }{2\lambda }\right)^2\right)
 \right].
\end{eqnarray}
Calculating the integral is trivial for the first term and is slightly more involved for the second one. We finally get:
\begin{eqnarray}\label{p-real-diff}
  \frac{d\mathcal{P}_{\mathbbm{R}^2}}{d\lambda d\bar{\lambda }}&=&
  \frac{\Delta ^3|\mathop{\mathrm{Im}}\lambda |}{2^{\frac{11}{2}}\pi ^{\frac{3}{2}}\sigma ^3|\lambda |^6}\,\,{\rm e}\,^{-\frac{\Delta ^2\left(\mathop{\mathrm{Re}}\lambda \right)^2}{8\sigma ^2|\lambda |^4}}
  K_0\left(\frac{\Delta ^2\left(\mathop{\mathrm{Im}}\lambda \right)^2}{8\sigma ^2|\lambda |^4}\right)
\nonumber \\
&&
  +\frac{\Delta }{2^\frac{5}{2}\pi ^{\frac{3}{2}}\sigma \lambda ^2}\,
  \,{\rm e}\,^{-\frac{\Delta ^2}{8\sigma ^2\lambda ^2}}
  {}_2F_2\left(1,1;\frac{1}{2}\,,\frac{3}{2}\,;\frac{\Delta ^2}{16\sigma ^2\lambda ^2}\right)\delta \left(\mathop{\mathrm{Im}}\lambda \right),
\end{eqnarray}
where ${}_2F_2$ is the hypergeometric function.

\begin{figure}[t]
\begin{center}
\subfigure[]{
   \includegraphics[height=4.1cm] {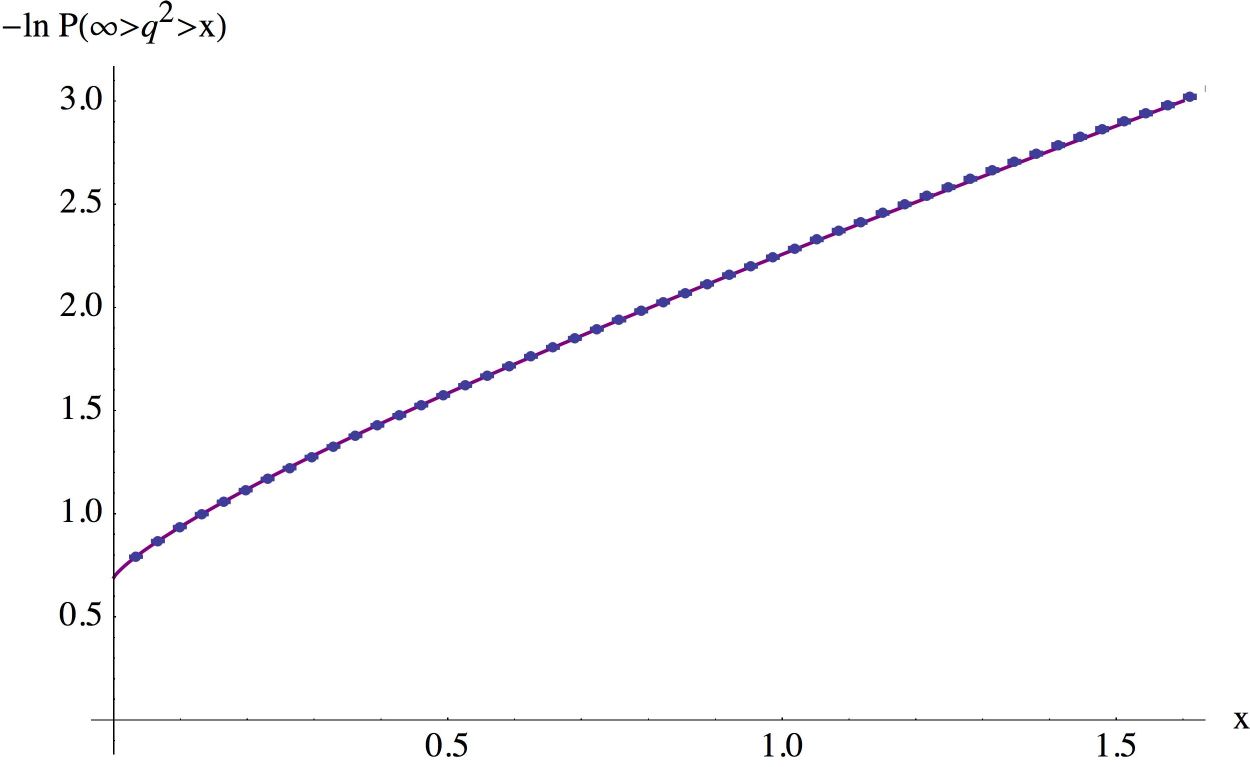}
   \label{fig2:subfig1-R}
 }
 \subfigure[]{
   \includegraphics[height=4.1cm] {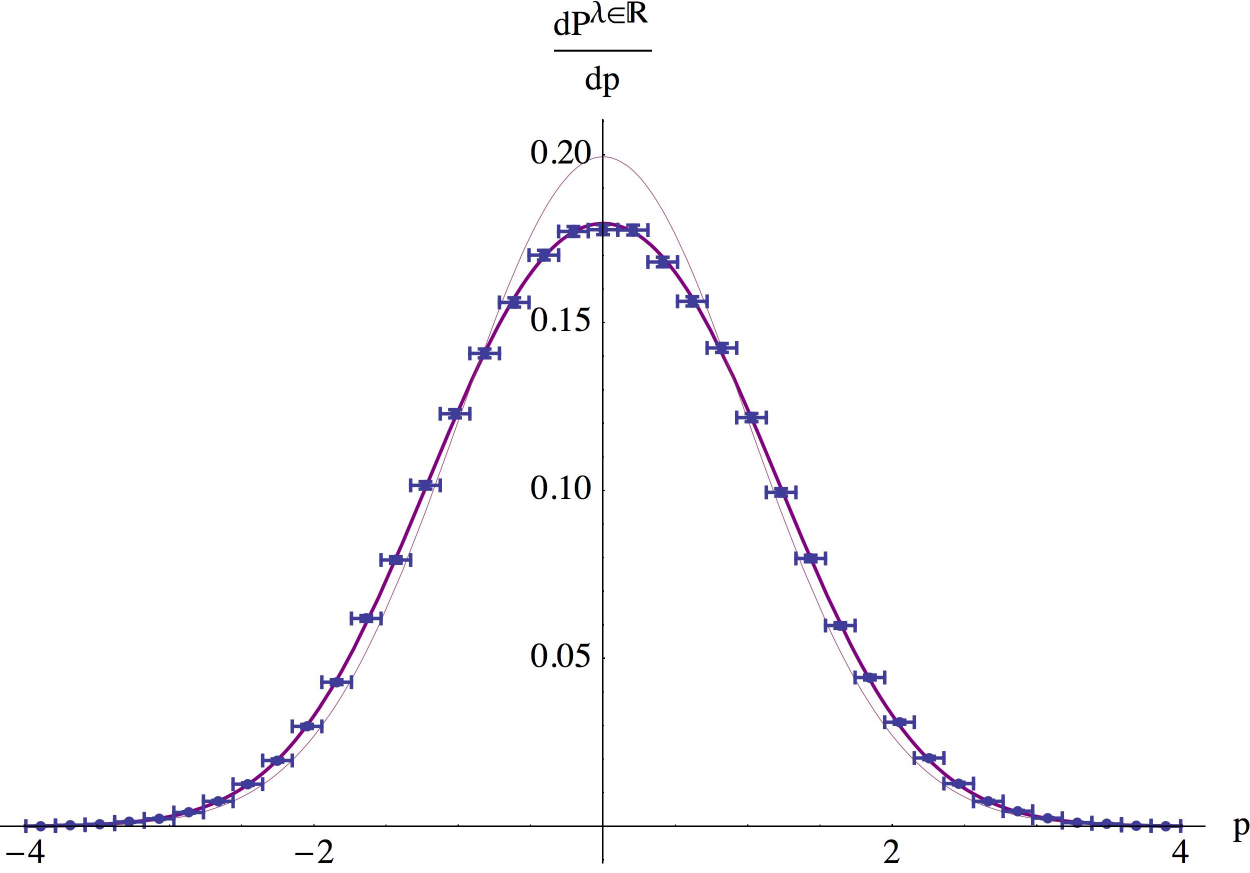}
   \label{fig2:subfig2-R}
 }
\caption{\label{Rlx}\small  The level-crossing probability for real $2\times 2$ matrices (\ref{rlsom}) with $H_0=\mathop{\mathrm{diag}}(1,-1)$ and $\sigma =1/\sqrt{2}$: (a) The cumulative distribution in the imaginary part of $1/\lambda $. (b) The differential distribution on the real axis. The thin line in the second plot is the Gaussian with the effective variance (\ref{sigma-eff}).
Blue dots are numerical data. Notice that in both cases the total integrated probability is $1/2$.}
\end{center}
\end{figure}

The level crossing for general real matrices is similar, in a way, to the case of Hermitian matrices considered before. The probability density is  not rotationally invariant, but  factorizes in the product of independent probabilities for the real and imaginary parts of $1/\lambda $. The natural variables are again $p$ and $q$ from~(\ref{p+iq}). 

The level-crossing points appear in the complex plane or on the real axis with equal
probability. Consider first the distribution  in the complex plane away from the real axis which is best characterized by the cumulative probability in $q^2$, for which we get:
\begin{equation}\label{rlsom}
 \mathcal{P}_{\mathbbm{R}^2}\left(\infty >q^2>\frac{8\sigma ^2x}{\Delta ^2}\right)
 =\frac{1-x\left(\mathbf{L}_{-1}(x)K_0(x)+\mathbf{L}_0(x)K_1(x)\right)}{2},
\end{equation}
where $\mathbf{L}_\nu (x)$ is the modified Struve function. The result is shown in fig.~\ref{fig2:subfig1-R}, where it is also compared to the numerical data. The total probability asymptotes to $1/2$ at large $x$. The other half of the level crossings happens on the real axis at $q^2=\infty $.

The probability distribution on the real axis is given by the second term in (\ref{p-real-diff}). In terms of the variable $p$ from (\ref{p+iq}),
\begin{equation}\label{realreal}
 \frac{d\mathcal{P}_{\mathbbm{R}^2}^{\lambda \in\mathbbm{R}}}{dp}
 =\frac{\Delta }{\left(2\pi \right)^{\frac{3}{2}}\sigma }\,
 \,{\rm e}\,^{-\frac{\Delta ^2p^2}{8\sigma ^2}}
 {}_2F_2\left(1,1;\frac{1}{2}\,,\frac{3}{2}\,;\frac{\Delta ^2p^2}{16\sigma ^2}\right).
\end{equation}
The probability, displayed in fig.~\ref{fig2:subfig2-R}, is very similar to a Gaussian with the effective variance
\begin{equation}\label{sigma-eff}
 \sigma ^2_{\rm eff}=\frac{8\sigma^2 }{\Delta^2 }\,,
\end{equation}
although it is somewhat flatter and more spead-out.

\subsection{Real matrices: general case}

Once we allow for non-Hermitian perturbations $V$, it is no longer natural to insist on the diagonal form of the initial matrix $H_0$. In all the cases considered before ($GUE$, $GOE$ and $GE_{\mathbbm{C}}$) this was not really a restriction. A generic Hermitian, real symmetric or complex matrix can be diagonalized by a unitary, orthogonal or $SL(N,\mathbbm{C})$ similarity transformation, respectively. These transformation are symmetries of the probability measures of $GUE$, $GOE$ and $GE_{\mathbbm{C}}$. But for $GE_{\mathbbm{R}}$ this is no longer true. Almost any real matrix can be diagonalized by an $SL(N,\mathbbm{R})$ transformation $H_0\rightarrow S^{-1}H_0S$, but this transformation is no longer a symmetry of the probability measure of  $GE_{\mathbbm{R}}$.

In this subsection we relax the condition that $H_0$ is Hermitian and allow $H_0$ to be generic but still fixed real matrix. It can  then be expanded as\footnote{We choose from the outset to deal with traceless matrices. The dependence on $\mathop{\mathrm{tr}}H_0$ drops out from the level-crossing PDF. To restore the full generality in the formulas below, $H_0$ should be replaced by $H_0-\mathbbm{1}\mathop{\mathrm{tr}}H_0/2$.}
\begin{equation}
 H_0=\varepsilon _1\sigma _1+i\varepsilon _2\sigma _2+\varepsilon _3\sigma _3.
\end{equation}
The level-crossing condition becomes
\begin{equation}\label{lcc}
 \left(\mathbf{v}+\frac{1}{\lambda }\,\boldsymbol{\varepsilon }\right)^2=0.
\end{equation}
Introducing, as before, the real and imaginary parts of $\lambda =x+iy$, we get two possible solutions that correspond to level crossings in the complex plane and on the real line:
\begin{eqnarray}\label{complexcr}
 {\rm Complex:}&\qquad &\left(\mathbf{v}+\frac{x}{|\lambda |^2}\,\boldsymbol{\varepsilon }\right)^2=\frac{y^2\ep^2}{|\lambda |^4}\,,\qquad 
 \left(\mathbf{v}+\frac{x}{|\lambda |^2}\,\ep\right)\cdot \ep=0.
 \\ 
 \label{realcr}
 {\rm Real:}&\qquad &y=0,\qquad  \left(\mathbf{v}+\frac{1}{\lambda  }\,\boldsymbol{\varepsilon }\right)^2=0.
\end{eqnarray}
The level-crossing probability, upon shifting the integration variable $\mathbf{v}\rightarrow \mathbf{v}-x\ep/|\lambda |^2$, becomes
\begin{eqnarray}\label{very-general-integral}
   \frac{d\mathcal{P}_{\mathbbm{R}^2}}{d\lambda d\bar{\lambda }}&=&
   \frac{\,{\rm e}\,^{-\frac{x^2\left(\ep,\ep\right)}{2\sigma ^2|\lambda |^4}}}{\left(2\pi \sigma ^2\right)^{\frac{3}{2}}}
   \int_{}^{}d^3v\,
   \,{\rm e}\,^{-\frac{\left(\mathbf{v},\mathbf{v}\right)}{2\sigma ^2}+\frac{x\left(\ep,\mathbf{v}\right)}{\sigma ^2|\lambda |^2}}
   \left[
   \frac{|y|\left(\ep^2\right)^2}{2|\lambda |^6}\,\delta \left(\ep\cdot \mathbf{v}\right)\delta \left(\mathbf{v}^2-\frac{y^2\ep^2}{|\lambda |^4}\right)
   \right.
\nonumber \\
   &&\left.
   +\frac{|\ep\cdot \mathbf{v}|}{2\lambda ^2}\,\delta \left(\mathbf{v}^2\right)\delta \left(y\right)
   \right].
\end{eqnarray}

The Lorentzian scalar product (\ref{Lorentz}) equips the space of three-vectors $\ep$ with  the usual causal structure of Special Relativity: time-like vectors for which the $SL(2,\mathbbm{R})$ invariant $\ep^2=\varepsilon _1^2-\varepsilon _2^2+\varepsilon _3^2$ is negative lie inside the light-cone $\ep^2=0$, while the exterior of the light-cone is formed by space-like vectors with the positive scalar square. The qualitative structure of level crossings crucially depends on whether the vector $\ep$ is time-like, space-like or null\footnote{The last case is degenerate and will not be considered in what follows.}. 

Complex level crossings are impossible when $\ep$ is time-like.   Indeed, any vector orthogonal to a time-like vector must be space-like. Hence, if $\ep$ is time-like, $\mathbf{v}+x\ep/|\lambda |^2$ has a positive scalar square,  in virtue of the second equation in (\ref{complexcr}), while the scalar square of $\ep$ is at the same time negative and the first equation thus has no solutions.

\subsubsection{Space-like $\ep$}

Consider first the case of space-like $\ep$, $\ep^2>0$. A particular example worked out in sec.~\ref{realms} corresponds to $\ep=\Delta \mathbf{e}_3/2$. We will use the same notation for the scalar square of $\ep$:
\begin{equation}
 \ep^2\equiv \frac{\Delta ^2}{4}\,.
\end{equation}
The $2\times 2$ matrix $H_0$ has a real spectrum for space-like $\ep$ and $\Delta $ so defined has the meaning of the gap between its two eigenvalues. Assuming that $H_0$ is traceless, the eigenvalues are $\pm \Delta /2$ and $\Delta ^2=-4\det H_0$.

It may seem that the level-crossing probability can only depend on the $SL(2)$ invariants. The unique such invariant associated with the matrix $H_0$ is $\Delta $.
But this assumption is not true. While the level-crossing condition (\ref{lcc}) is expressed in terms of the Minkowski scalar product, and can indeed be brought to the form (\ref{vgrill}) by a Lorentz transformation, the probability measure depends on the Euclidean scalar products and is not Lorentz-invariant. The level-crossing probability therefore will depend on the additional parameters of the matrix $H_0$ (equivalently,  of vector $\ep$), which are not $SL(2)$ invariants.

\begin{figure}[t]
\begin{center}
\subfigure[]{
   \includegraphics[width=5cm] {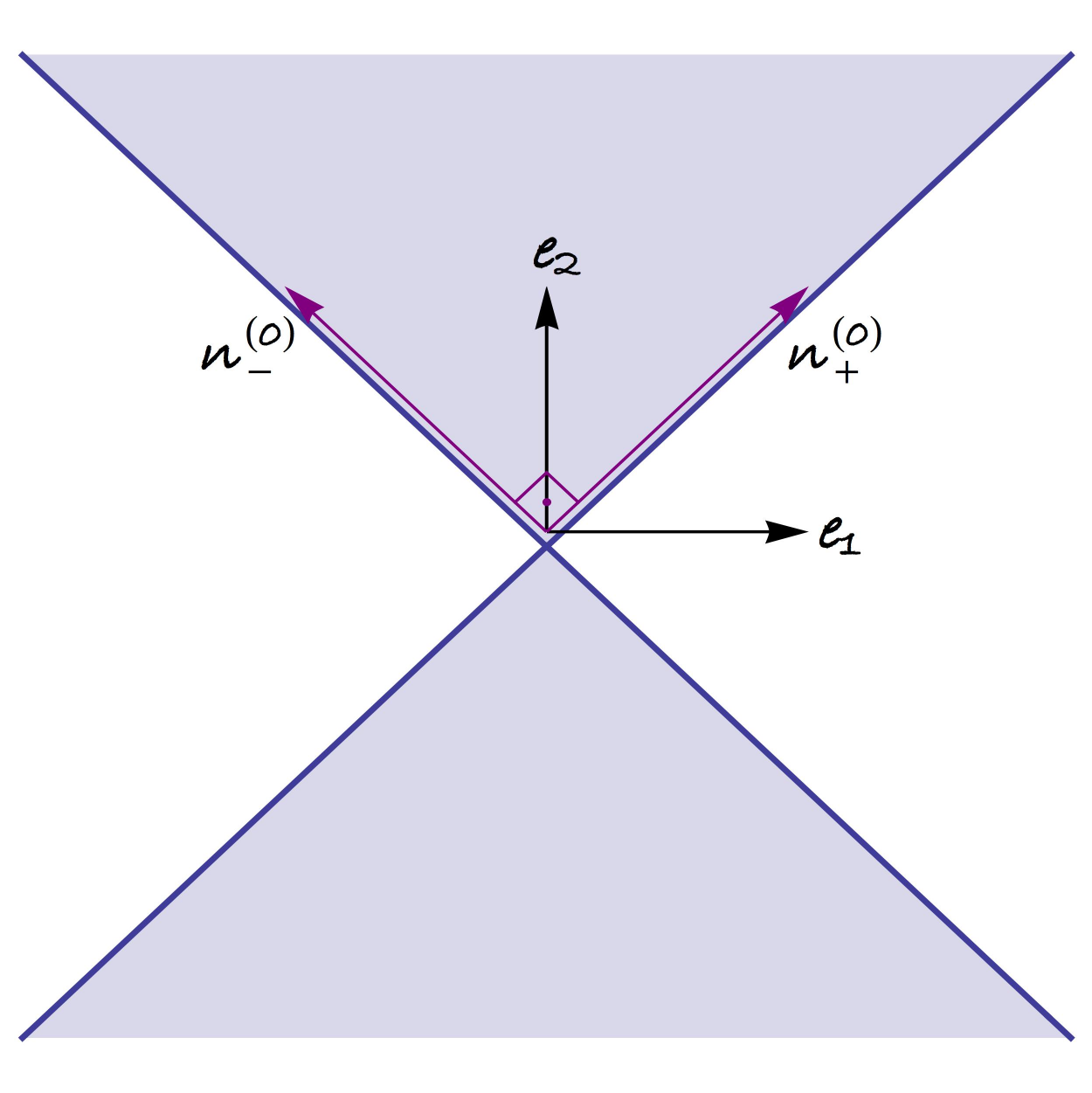}
   \label{fig2:subfig1-LC}
 }
 \subfigure[]{
   \includegraphics[width=5cm] {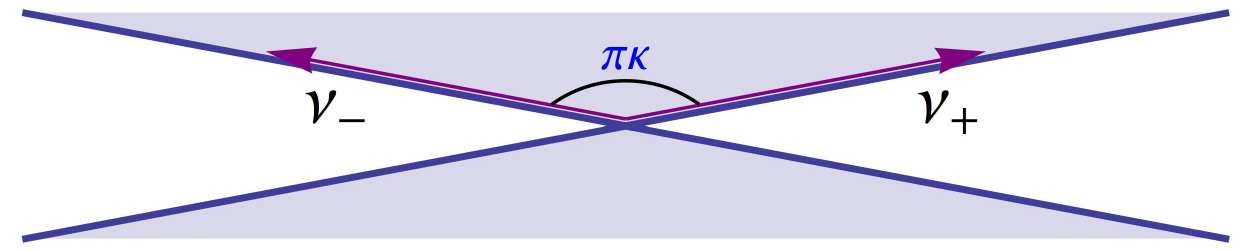}
   \label{fig2:subfig2-LC}
 }
\caption{\label{Lcn}\small  The shaded area is the space of solutions for the real level-crossing condition: (a) The cross section by the $(12)$ plane for the case of diagonal $H_0$. (b) The cross section perpendicular to $\et_3$ in the general case. $\kappa $ is the fraction of real level crossings.}
\end{center}
\end{figure}

To illustrate the point, let us calculate the fraction of real level crossings as a function of $\ep$:
\begin{equation}
 \kappa =\mathcal{P}_{\mathbbm{R}^2}\left(\mathop{\mathrm{Im}}\lambda = 0\right)=\left\langle 
 \frac{\#_{{\rm Real~Level-Crossings}}}{\#_{{\rm Level-Crossings}}} 
 \right\rangle
\end{equation}
For a diagonal $H_0$, complex and real crossings happen with equal probability, corresponding to $\kappa =1/2$. This fact has a simple geometric interpretation. For a given $\lambda $, the solutions of the real level-crossing condition (\ref{vgrill})  form the light-cone centered at the point $\Delta \mathbf{e}_3/2\lambda $. As $\lambda $ varies, the solutions fill the space between the two light-sheets
\begin{equation}\label{light-sheets}
 \mathcal{L}_\pm^{(0)}:~\alpha \mathbf{e}_3+\beta \mathbf{n}_\pm^{(0)},
 \qquad \alpha ,\beta \in\left(-\infty ,+\infty \right),
\end{equation}
where $\mathbf{n}_\pm^{(0)}$ are the null vectors perpendicular to $\mathbf{e}_3$:
\begin{equation}
 \mathbf{n}_\pm^{(0)}=\mathbf{e}_2\pm\mathbf{e}_1.
\end{equation}
The cross section of this space by the $(12)$ plane is shown in fig.~\ref{fig2:subfig1-LC}. The fraction of real level crossings is the volume of the space of solutions with respect to the Gaussian probability measure. Since the measure is rotationally invariant, the fraction of real eigenvalues is simply the relative proportion of the shaded area in the figure, which is exactly $1/2$.

\begin{figure}[t]
\begin{center}
 \centerline{\includegraphics[width=4.2cm]{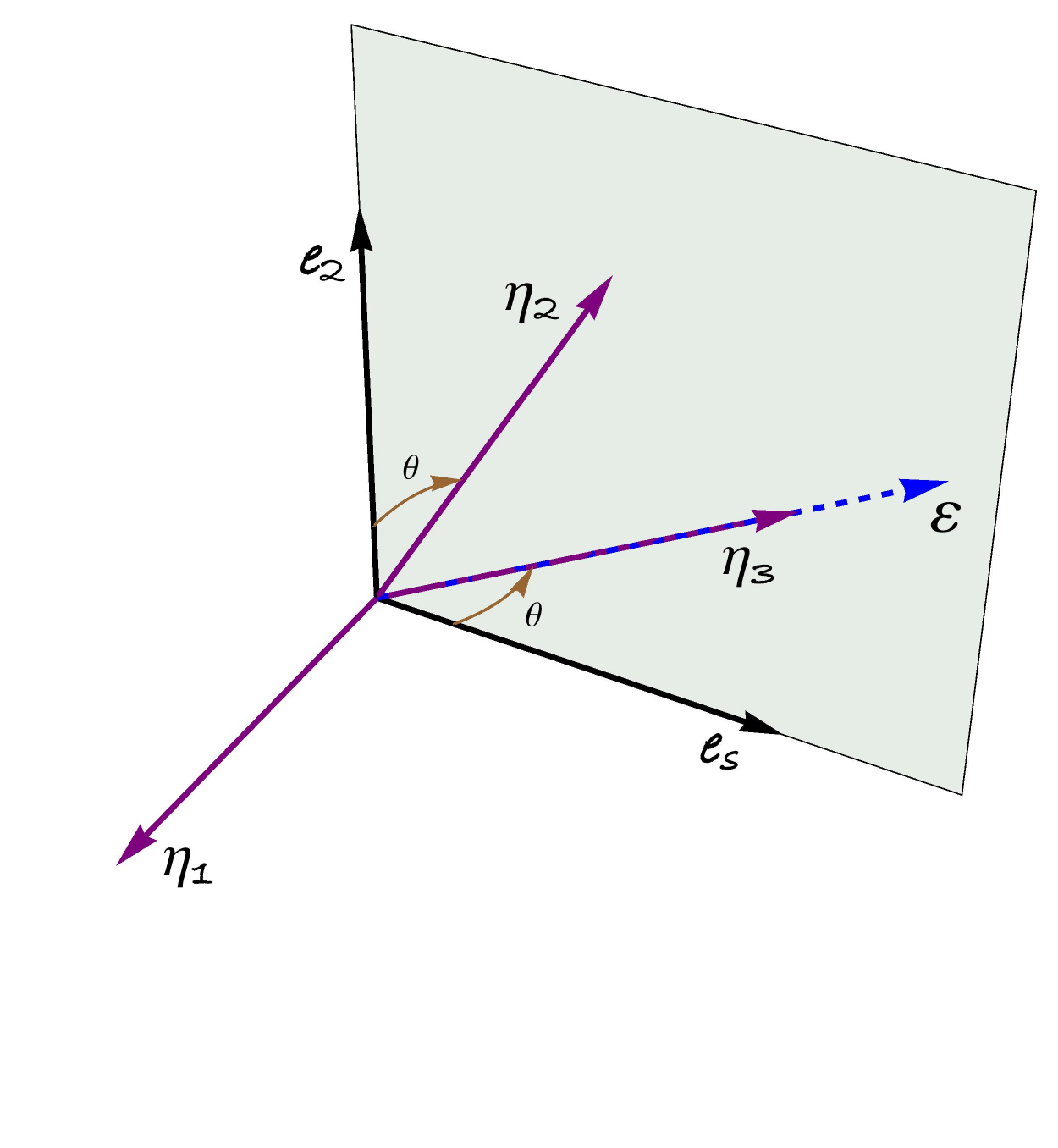}}
\caption{\label{loc-frame}\small The orthonomal basis in the rest frame of $\ep$.}
\end{center}
\end{figure}

To generalize this argument to arbitrary $\ep$, we can introduce the orthonormal basis $\et_\mu $, obtained by boosting $\mathbf{e}_\mu $ to the rest frame of $\ep$, such that
\begin{equation}
 \ep=\frac{\Delta }{2}\,\et_3
\end{equation}
and
\begin{equation}\label{canbas}
 \et_\mu \cdot \et_\nu =\mathop{\mathrm{diag}}\left(1,-1,1\right).
\end{equation}
The second vector, $\et_{2}$, is obtained by a Lorentz transformation from $\mathbf{e}_2$ and $\mathbf{e}_s$, where $\mathbf{e}_s$ is the unit vector along the intersection of the $(13)$ and $(\ep,\mathbf{e}_2)$ planes (fig.~\ref{loc-frame}):
\begin{eqnarray}\label{etwo}
 \mathbf{e}_2&=&\et_2\cosh{\theta }-\et_3\sinh{\theta }
\\
\label{ethree}
\mathbf{e}_s&=&\et_3\cosh{\theta }-\et_2\sinh{\theta }\,.
\end{eqnarray}
The first equation defines $\et_2$ in terms of $\et_3$ and $\mathbf{e}_2$. The second one can be taken as a definition of $\mathbf{e}_s$. Finally, one can take $\et_1=\mathbf{e}_2\times \mathbf{e}_s$.

The rapidity $\theta $ can be found from (\ref{etwo}):
\begin{equation}\label{rapidity}
 \sinh{\theta }=-\et_3\cdot \mathbf{e}_2=\frac{2\varepsilon _2}{\Delta }
 =\frac{\varepsilon _2}{\sqrt{\varepsilon _1^2-\varepsilon _2^2+\varepsilon _3^2}}=\sqrt{-\frac{\det\left(H_0-H_0^t\right)}{4\det H_0}}\,,
\end{equation}
and takes arbitrary positive values. The case of $\theta =0$ corresponds to the setup  of sec.~\ref{realms}, when the matrix $H_0$ is real symmetric.
The rapidity is the other parameter, in addition to $\Delta $, on which the level-crossing PDF will depend.
This happens because the basis  $\et_\mu $,  orthonormal with respect to the Minkowski scalar product,  is not canonically normalized with respect to the Euclidean scalar product. From (\ref{etwo}), (\ref{ethree}) we find:
\begin{equation}\label{newmetric}
 \left(\et_\mu ,\et_\nu \right)=\begin{pmatrix}
  1 & 0 & 0 \\ 
   0 & \cosh 2\theta  & \sinh 2\theta  \\ 
   0 & \sinh 2\theta  & \cosh 2\theta  \\ 
 \end{pmatrix}.
\end{equation}

The equation for the light-sheets (\ref{light-sheets}) for arbitrary $\ep$ becomes
\begin{equation}\label{lis}
 \mathcal{L}_\pm:~\alpha \et_3+\beta \mathbf{n}_\pm,
 \qquad \alpha ,\beta \in\left(-\infty ,+\infty \right)
\end{equation}
with
\begin{equation}
 \mathbf{n}_\pm=\et_2\pm\et_1.
\end{equation}
We can again exploit the rotational symmetry of the probability density, now with respect to rotations around the $\et_3$-axis. To find the fraction of real level crossings, we need to disect the light-sheets (\ref{lis}) by the plane passing through the origin and perpendicular (in the Euclidean metric) to $\et_3$, in other words to find two vectors $\nd_\pm\in\mathcal{L}_\pm$such that $(\nd_\pm,\et_3)=0$. The fraction of real level crossings is given by the angle between these two vectors, as should be clear from fig.~\ref{fig2:subfig2-LC}:
\begin{equation}
 \cos\pi \kappa =\frac{\left(\nd_+,\nd_-\right)}{\sqrt{\left(\nd_+,\nd_+\right)\left(\nd_-,\nd_-\right)}}\,.
\end{equation}
With the help of (\ref{newmetric}) we find:
\begin{equation}
 \nd_\pm=\mathbf{n}_\pm-\et_3\tanh 2\theta, 
\end{equation}
and consequently
\begin{equation}\label{cospikappa}
 \cos\pi\kappa =-\tanh^2{\theta }\,,
\end{equation}
so that in general $\kappa \geq 1/2$. Using the explicit expression for the rapidity, the fraction of real level crossings can be rewritten as
\begin{equation}\label{coskappa}
 \cos\pi \kappa =-\frac{\varepsilon _2^2}{\varepsilon _1^2+\varepsilon _3^2}=
 \frac{\det\left(H_0-H_0^t\right)}{\det\left(H_0+H_0^t\right)}\,.
\end{equation}

\begin{figure}[t]
\begin{center}
 \centerline{\includegraphics[width=7cm]{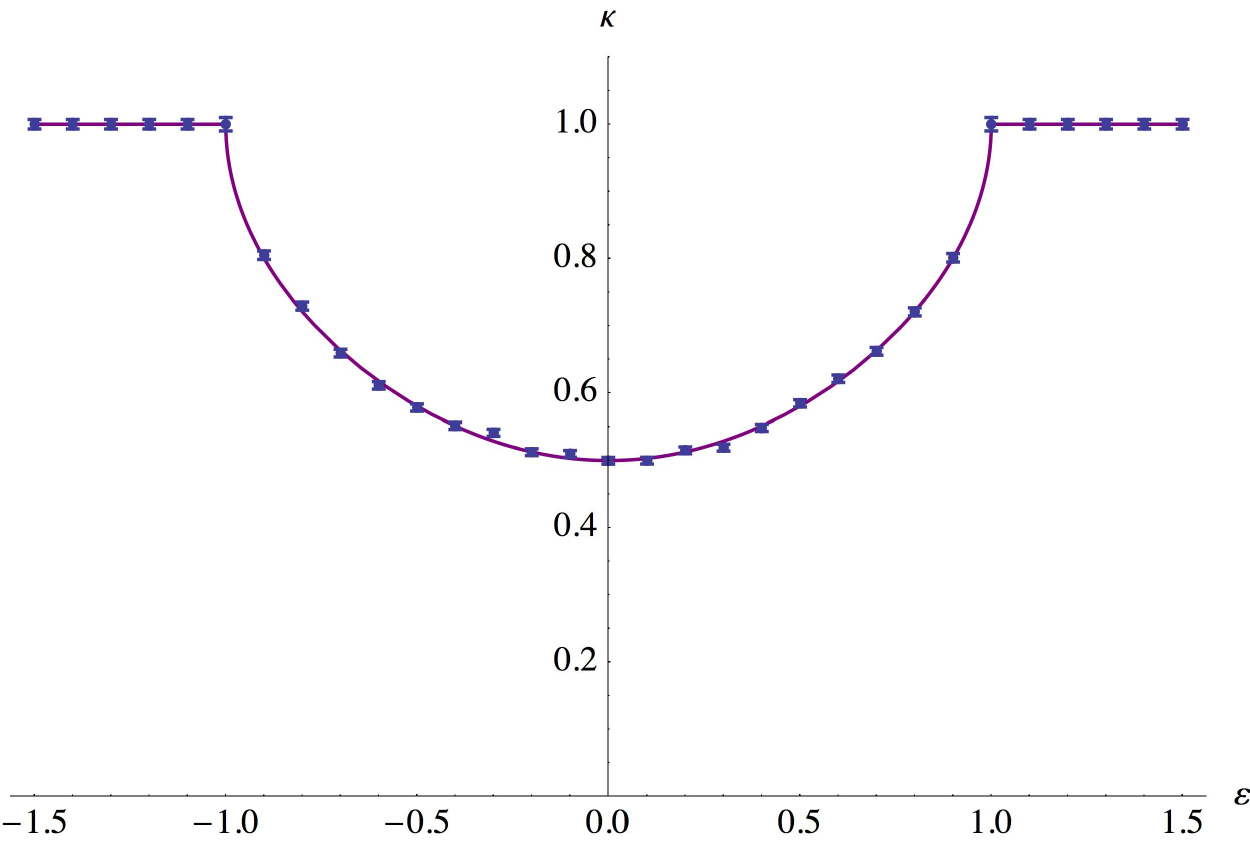}}
\caption{\label{real-fraction}\small The fraction of real level crossings for $H_0$ given by (\ref{sampleH0}), given by eq.~(\ref{samplekappa}). The dots represent numerical data.}
\end{center}
\end{figure}

In fig.~\ref{real-fraction} the fraction of real level crossings is plotted as a function $\varepsilon $ for the matrix of the form
\begin{equation}\label{sampleH0}
 H_0=\begin{pmatrix}
  1 & \varepsilon  \\ 
  -\varepsilon  & -1  \\ 
 \end{pmatrix}.
\end{equation}
 In this case (\ref{coskappa}) gives 
\begin{equation}\label{samplekappa}
 \kappa =1-\frac{\arccos\varepsilon ^2}{\pi }\,.
\end{equation}

The level-crossing PDF is given by the integral (\ref{very-general-integral}). It is convenient to expand the integration variable in $\et_\mu $: $\mathbf{v}=v_\mu \et_\mu $, and use (\ref{newmetric}) to express the scalar products in the probability measure in terms of $v_\mu $ and the rapidity $\theta $ given by (\ref{rapidity}).
The delta-functions eliminate two integrations for the complex level crossings and one integration for the real ones. In the latter case, one more integration can be performed with the help of  the change of variables $v_2=v\cosh\eta $, $v_3=v\sinh\eta $. Altogether we get:
\begin{eqnarray}\label{awful mess}
  \frac{d\mathcal{P}_{\mathbbm{R}^2}}{d\lambda d\bar{\lambda }}&=&
  \frac{\Delta ^2}{16\pi ^{\frac{3}{2}}\sigma ^2|\lambda |^4}\,
  \mathcal{F}\left(\frac{\Delta \mathop{\mathrm{Re}}\lambda }{2\sqrt{2}\sigma |\lambda |^2}\,,\frac{\Delta \mathop{\mathrm{Im}}\lambda }{2\sqrt{2}\sigma |\lambda |^2}\,;\theta \right)
\nonumber \\
&&
  +
  \frac{\Delta }{2^{\frac{5}{2}}\pi ^{\frac{3}{2}}\sigma \lambda ^2}\,
  \mathcal{G}\left(\frac{\Delta }{4\sigma \lambda }\,;\theta \right)
  \delta \left(\mathop{\mathrm{Im}}\lambda \right),
\end{eqnarray}
where the first term described complex level crossings and the second term - real ones. The functions $\mathcal{F}$ and $\mathcal{G}$ are given by
\begin{eqnarray}
 \mathcal{F}(z,w;\theta )&=&|w|\,{\rm e}\,^{-
 z^2\cosh 2\theta -w^2
 }
 \int_{-\infty }^{+\infty }\frac{dv}{\sqrt{v^2+w^2}}\,\,
 \,{\rm e}\,^{-v^2\left(\cosh 2\theta +1\right)+2vz\sinh 2\theta }
\nonumber \\
\mathcal{G}(z;\theta )&=&\,{\rm e}\,^{-2z^2\cosh 2\theta }\left(
 1+\sinh\theta \arctan\sinh\theta 
 \vphantom{\frac{\sqrt{\pi }z}{2}\int_{-\infty }^{+\infty }
 d\eta \,|\sinh\eta |\,
 \frac{\sinh(\eta +2\theta )}{\cosh^3(\eta +\theta )}\,
 \mathop{\mathrm{erf}}\left(z\,\frac{\sinh(\eta +2\theta )}{\cosh(\eta +\theta )}\right)
 \,{\rm e}\,^{z^2\,\frac{\sinh^2(\eta +2\theta )}{\cosh^2(\eta +\theta )}}}
 \right.
\nonumber \\
&&\left.
 +\frac{\sqrt{\pi }z}{2}\int_{-\infty }^{+\infty }
 d\eta \,|\sinh\eta |\,
 \frac{\sinh(\eta +2\theta )}{\cosh^3(\eta +\theta )}\,
 \mathop{\mathrm{erf}}\left(z\,\frac{\sinh(\eta +2\theta )}{\cosh(\eta +\theta )}\right)
  \right.
\nonumber \\
&&\times \left.
 \,{\rm e}\,^{z^2\,\frac{\sinh^2(\eta +2\theta )}{\cosh^2(\eta +\theta )}}
\right).
\end{eqnarray}
One can check that at $\theta =0$ eq.~(\ref{awful mess}) reduces to (\ref{p-real-diff}).

Writing $1/\lambda =p+iq$, we find for the cumulative distribution along the imaginary axis:
\begin{equation}\label{rlsom1}
 \mathcal{P}_{\mathbbm{R}^2}\left(\infty >q^2>\frac{8\sigma ^2x}{\Delta ^2}\right)
 =
 \frac{1}{\pi \sqrt{\cosh 2\theta }}\int_{x}^{\infty }d\rho \,\,{\rm e}\,^{-\frac{\cosh 2\theta -1}{2\cosh 2\theta }\,\rho }K_0\left(\frac{\cosh 2\theta +1}{2\cosh 2\theta }\,\rho \right).
\end{equation}
This equation generalizes (\ref{rlsom}).
In particular, the total fraction of complex level crossings is
\begin{equation}
\mathcal{P}_{\mathbbm{R}^2}\left(\infty >q^2>0\right)=\frac{1}{\pi }\,
\arccos(\tanh^2\theta),
\end{equation}
in agreement with (\ref{cospikappa}). The probability distribution of level crossings on the real axis is
\begin{equation}\label{genrealreal}
 \frac{d\mathcal{P}_{\mathbbm{R}^2}^{\lambda \in\mathbbm{R}}}{dp}
 =\frac{\Delta }{\left(2\pi \right)^{\frac{3}{2}}\sigma }\,
 \mathcal{G}\left(\frac{\Delta p}{4\sigma }\,;\theta \right),
\end{equation}
which generalizes (\ref{realreal}) to $\theta \neq 0$.

\begin{figure}[t]
\begin{center}
\subfigure[]{
   \includegraphics[height=4.1cm] {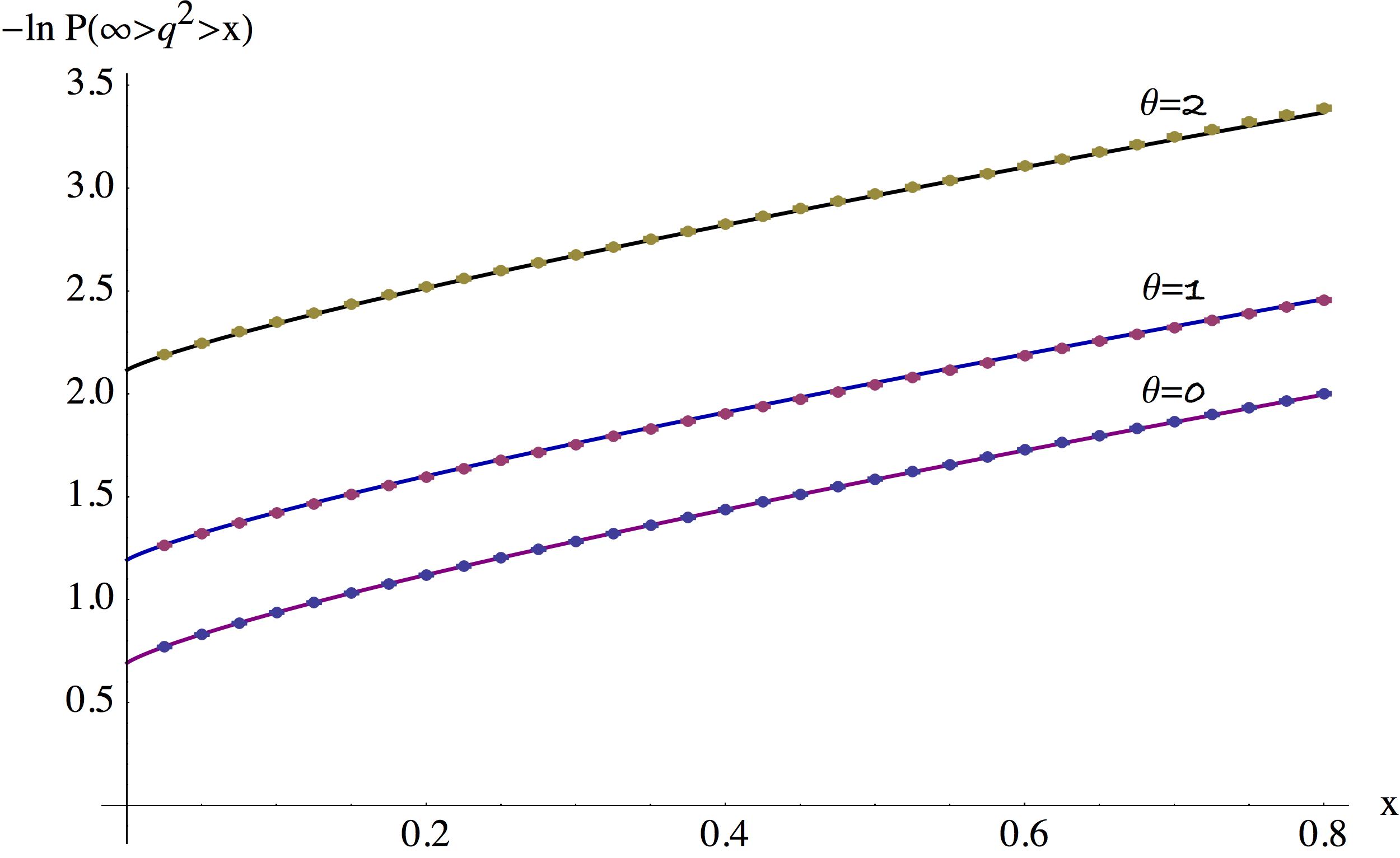}
   \label{fiR-gen-re}
 }
 \subfigure[]{
   \includegraphics[height=4.1cm] {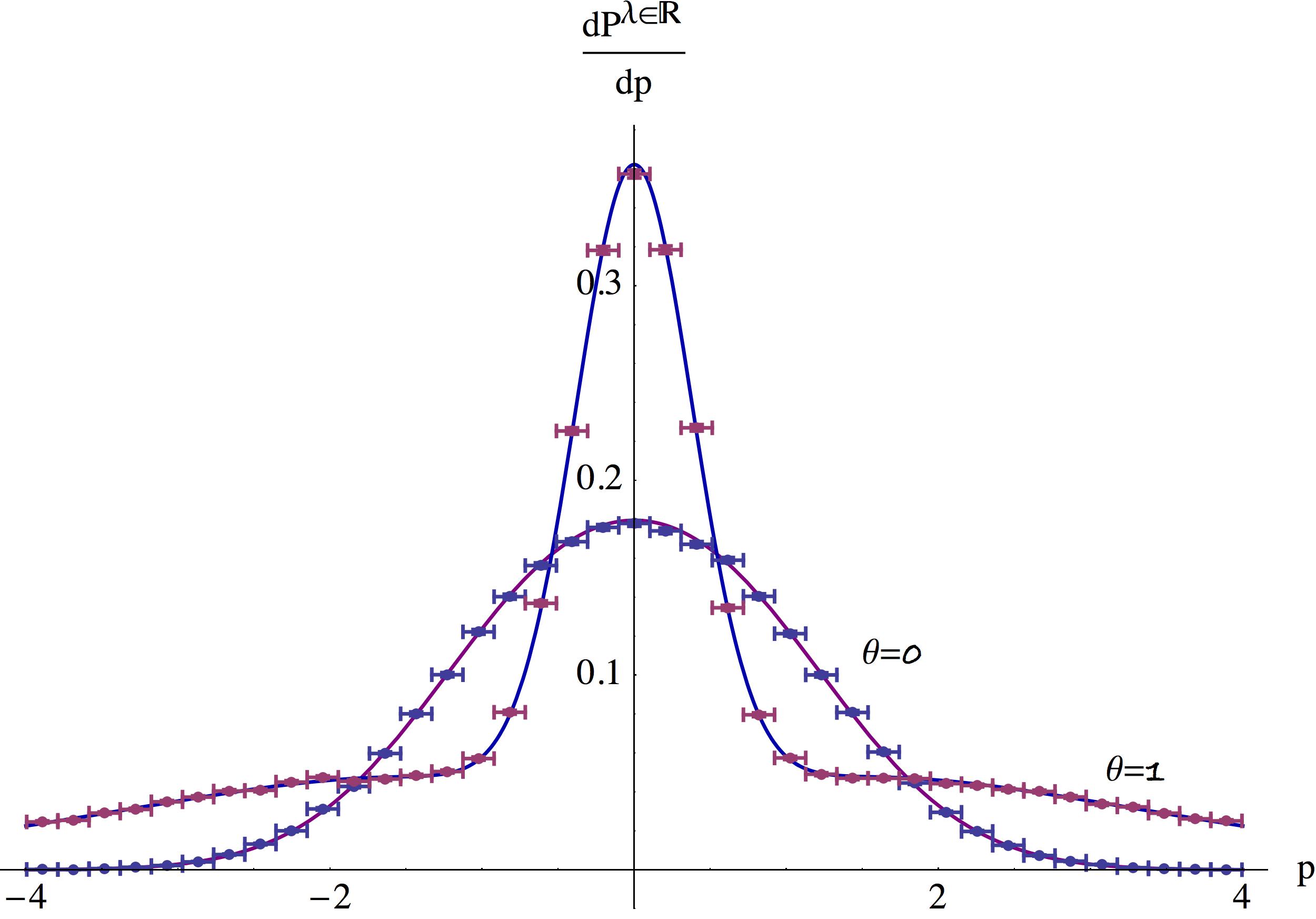}
   \label{fiR-gen-im}
 }
\caption{\label{Re-gen}\small  The level-crossing probability for general real matrices in the space-like case: (a) Cumulative probability along the imaginary axis (\ref{rlsom1}) for $\theta =0,1,2$. (b) Differential probability on the real axis (\ref{genrealreal}) for $\theta =0$ and $1$. In all cases $\Delta =2$ and $\sigma =1/\sqrt{2}$.}
\end{center}
\end{figure}
These results are illustrated in fig.~\ref{Re-gen}. The probability of complex level crossings is very well approximated by the Poisson distribution in $q^2$, appropriately normalized. The probability of real level crossings has more structure. While at $\theta =0$ the distribution is very similar to  Gaussian, the probability density develops a sharp peak at zero at larger $\theta $ and at the same time has a longer tail that very slowly relaxes to zero.

\subsubsection{Timelike $\ep$}

When $\ep^2<0$ the two conditions in (\ref{complexcr}) are incompatible and all the level crossings occur on the real line, according to (\ref{realcr}). The matrix $H_0$ now has two complex eigenvalues separated by  $i\tilde{\Delta }$ where
\begin{equation}
 \ep^2=-\frac{\tilde{\Delta }^2}{4}\,.
\end{equation}
As before we introduce the unit-norm vector along $\ep$:
\begin{equation}
 \et_2=\frac{2}{\tilde{\Delta }}\,\ep,
\end{equation}
which is now timelike: $\et_2^2=-1$, and define  the canonically normalized basis (\ref{canbas}) by a Lorentz transformation
\begin{eqnarray}\label{ltrtilde}
 \mathbf{e}_2&=&\et_2\cosh\tilde{\theta }-\et_3\sinh\tilde{\theta }
\nonumber \\
 \mathbf{e}_s&=&\et_3\cosh\tilde{\theta }-\et_2\sinh\tilde{\theta },
\end{eqnarray}
with the rapidity is given by
\begin{equation}
 \cosh\tilde{\theta }=-\et_2\cdot \mathbf{e}_2=\frac{2\varepsilon _2}{\tilde{\Delta }}=\frac{\varepsilon _2}{\sqrt{\varepsilon _2^2-\varepsilon _1^2-\varepsilon _3^2}}
 =\sqrt{\frac{\det\left(H_0-H_0^t\right)}{4\det H_0}}\,.
\end{equation}
The first equation in (\ref{ltrtilde}) defines $\et_3$, the second then determines $\mathbf{e}_s$ and we can take $\et_1=\mathbf{e}_2\times \mathbf{e}_s$. Since the Lorentz transformation (\ref{ltrtilde}) has the same form as (\ref{etwo}), (\ref{ethree}), the metric $(\et_\mu ,\et_\mu )$ is given by (\ref{newmetric}), up to the replacement $\theta \rightarrow \tilde{\theta }$.

\begin{figure}[t]
\begin{center}
 \centerline{\includegraphics[width=8cm]{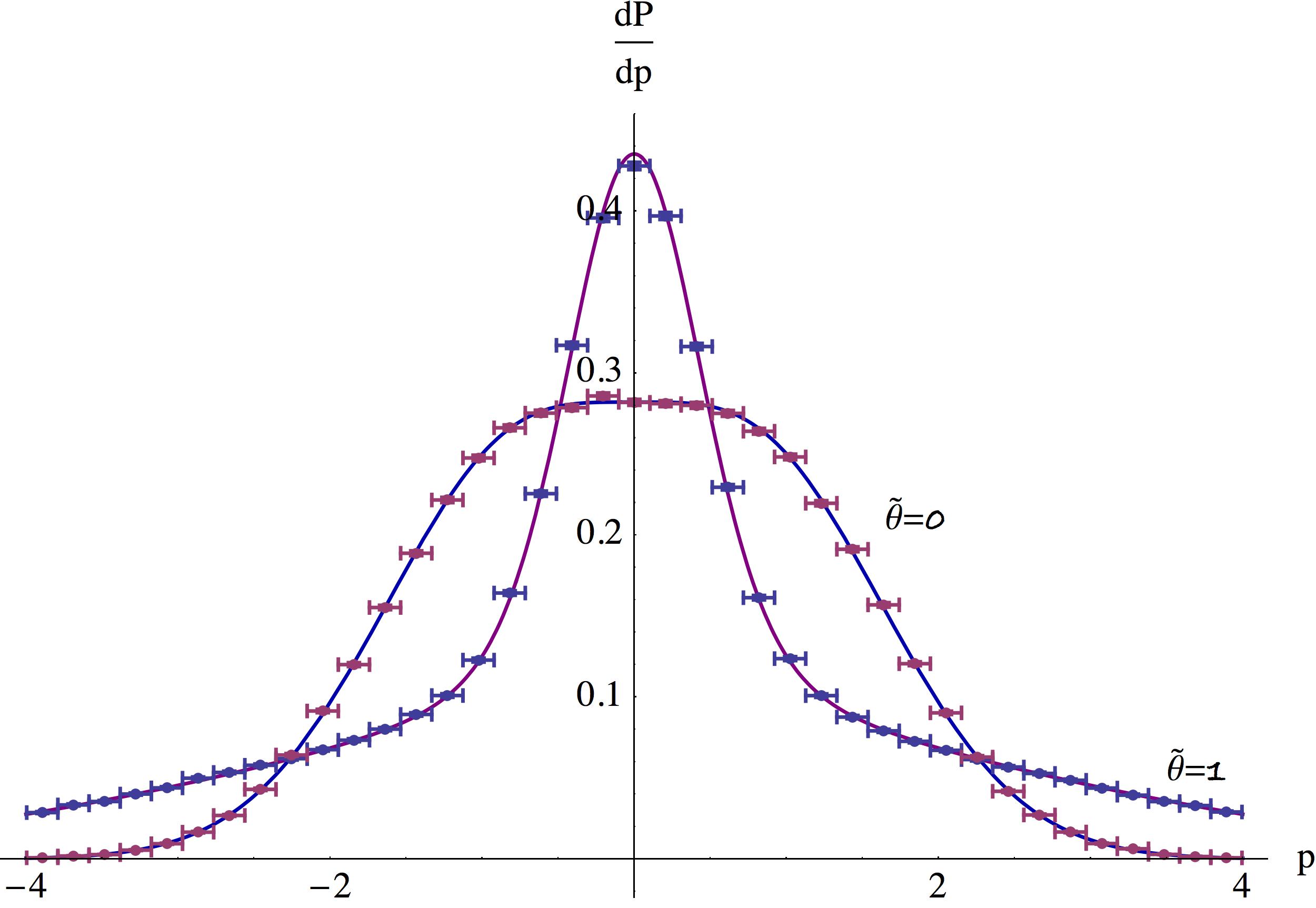}}
\caption{\label{real-timelike}\small The PDF (\ref{genrealrealtilde}) for time-like real matrices, for $\tilde{\theta }=1$ and $0$, $\Delta =2$ and $\sigma =1/\sqrt{2}$.}
\end{center}
\end{figure}

Expanding the integration variable in (\ref{very-general-integral}) in the basis of $\et_\mu $: $\mathbf{v}=v_\mu \et_\mu $, eliminating $v_1$ via the delta-function and changing variables as $v_2=v\cosh\eta $, $v_3=v\sinh\eta $, we find for the PDF on the real axis of $p=1/\lambda $:
\begin{equation}\label{genrealrealtilde}
 \frac{d\mathcal{P}_{\mathbbm{R}^2}}{dp}
 =\frac{\tilde{\Delta} }{\left(2\pi \right)^{\frac{3}{2}}\sigma }\,
 \tilde{\mathcal{G}}\left(\frac{\tilde{\Delta} p}{4\sigma }\,;\tilde{\theta }\right),
\end{equation}
where
\begin{eqnarray}
\tilde{\mathcal{G}}(z;\theta )&=&\,{\rm e}\,^{-2z^2\cosh 2\theta }\left(
 \frac{\pi }{2}\cosh\theta 
 \vphantom{\frac{\sqrt{\pi }z}{2}\int_{-\infty }^{+\infty }
 d\eta \,\cosh\eta \,
 \frac{\cosh(\eta +2\theta )}{\cosh^3(\eta +\theta )}\,
 \mathop{\mathrm{erf}}\left(z\,\frac{\sinh(\eta +2\theta )}{\cosh(\eta +\theta )}\right)
 \,{\rm e}\,^{z^2\,\frac{\sinh^2(\eta +2\theta )}{\cosh^2(\eta +\theta )}}}
 \right.
\nonumber \\
&&\left.
+ \frac{\sqrt{\pi }z}{2}\int_{-\infty }^{+\infty }
 d\eta \,\cosh\eta \,
 \frac{\cosh(\eta +2\theta )}{\cosh^3(\eta +\theta )}\,
 \mathop{\mathrm{erf}}\left(z\,\frac{\cosh(\eta +2\theta )}{\cosh(\eta +\theta )}\right)
  \right.
\nonumber \\
&&\times \left.
 \,{\rm e}\,^{z^2\,\frac{\cosh^2(\eta +2\theta )}{\cosh^2(\eta +\theta )}}
\right).
\end{eqnarray}
It is straightforward to check that the PDF given by these formulas is normalized to one. The probability distribution is displayed in fig.~\ref{real-timelike}.

\section{Summary}

We studied probability distributions of level-crossing points for various ensembles of random matrices. The results depend on the ensemble at hand and on the matrix size, but some universal features do emerge from our analysis. First of all, there are certain similarities between $GUE_2$ and $GE^{\mathbbm{R}}_2$ ensembles, where the distribution factorizes into two independent distributions for the real and imaginary parts of the coupling parameter. There is also a similarity between $GOE_2$ and $GE^{\mathbbm{C}}_2$, in which case the distribution is rotationally invariant. While for complex matrices invariance under rotations follows from the intrinsic symmetries of the random matrix ensemble,  the phase independence of the PDF for real symmetric matrices comes as a surprise. The rotational symmetry for $2\times 2$ matrices follows from the explicit calculation \cite{zirnbauer1983destruction}. We checked numerically that rotational invariance persists for matrices of a larger size, but we could not explain this result by any obvious symmetry. We formulate this statement as the following conjecture.

{\bf Conjecture.} The level-crossing probability density for $GOE_N$ (real symmetric $N\times N$ matrices) $d\mathcal{P_{O_N}}(\lambda ,\bar{\lambda })/d\lambda d\bar{\lambda }$ is invariant under $\lambda \rightarrow \,{\rm e}\,^{i\alpha }\lambda $, $\bar{\lambda }\rightarrow \,{\rm e}\,^{-i\alpha }\bar{\lambda }$, and depends only on $|\lambda |$.

It would be interesting to study a more general setup where the initial matrix, which we have currently fixed, is also allowed to fluctuate. We plan to return to this problem in the near future.

\subsection*{Acknowledgements}

We would like to thank L.~Pastur and J.~Verbaarschot for dicussions.
B.S. wants to thank the Department of Mathematics of UIUC for the hospitality in June-July 2015 when a part of this project was carried out. 
The research of the K.Z. was supported by the Marie
Curie network GATIS of the European Union's FP7 Programme under REA Grant
Agreement No 317089, by the ERC advanced grant No 341222, by the Swedish Research Council (VR) grant
2013-4329, and by RFBR grant 15-01-99504. Finally, we are grateful to anonymous referees for their constructive criticism which allowed us to improve the quality of the exposition.  
 
\appendix

\bibliographystyle{nb}

\end{document}